\documentclass[11pt]{article}%

\usepackage[a4paper,margin=2.7cm]{geometry}
\usepackage[latin1]{inputenc}
\usepackage{amsmath}
\usepackage{amsfonts}
\usepackage{amssymb}
\usepackage{listings}
\usepackage{graphicx}
\usepackage{epstopdf}

\providecommand{\U}[1]{\protect\rule{.1in}{.1in}}

\begin{document}

\title{A Java library to perform S-expansions \\of Lie algebras}
\author{C. Inostroza$^{1}$\thanks{c.inostroza@gmail.com}, I. Kondrashuk$^{2}$\thanks{igor.kondrashuk@ubiobio.cl}, N. Merino$^{3}$\thanks{merinomo@apc.in2p3.fr} and F.
Nadal$^{4}$\thanks{felip.nadal@gmail.com }\\$^{1}${\small Departamento de Física, Universidad de Concepción,}\\{\small Casilla 160-C, Concepción, Chile}\\$^{2}${\small Grupo de Matem\'atica Aplicada, Departamento de Ciencias
Básicas, }\\{\small Univerdidad del Bío-Bío, Campus Fernando May, Casilla 447, Chillán,
Chile}\\$^{3}${\small APC, Universite Paris Diderot, }\\{\small 10, rue Alice Domon et Leonie Duquet, 75205 Paris Cedex 13, France.}\\$^{4}${\small Instituto de Física Corpuscular (IFIC), Edificio Institutos de
Investigación. }\\{\small c/ Catedrático José Beltrán, 2. E-46980 Paterna. España.}}
\maketitle

\begin{abstract}
The contraction method is a procedure that allows to establish non-trivial relations between Lie algebras and has had succesful applications in both mathematics and theoretical physics. This work deals with generalizations of the contraction procedure with a main focus in the so called S-expansion method as it includes most of the other generalized contractions. Basically, the S-exansion combines a Lie algebra $\mathcal{G}$ with a finite abelian semigroup $S$ in order to define new S-expanded algebras.
After giving a description of the main ingredients used in this paper, we present a Java library that automatizes the S-expansion procedure. 
With this computational tool we are able to represent Lie algebras and semigroups, so we can perform S-expansions of Lie algebras using arbitrary semigroups. 
We explain how the library methods has been constructed and how they work; then we give a set of example programs aimed to solve different problems. 
They are presented so that any user can easily modify them to perform his own calculations, without being necessarily an expert in Java. Finally, some comments about further developements and possible new applications are made.

\end{abstract}

\newpage\newpage

\tableofcontents

\newpage\newpage

\section{Introduction}

The theory of Lie groups and algebras plays an important role in physics, as it allows to describe the continues symmetries of a given physical system.
Due to the well-known relation between symmetries and conservation laws, via the Noether theorem, the Lie theory represents an essential ingredient in the construction of quantum field theories, particularly in the Standard Model of particles, as well as in general relativity and its generalizations.
Since the second half of the last century, the study of non-trivial\footnote{By ``non-trivial relations'' we mean that starting from a given algebra there are mechanisms allowing us to generate new algebras that are not isomorphic to the original one, i.e., they cannot be obtained by a simple change of basis.} relations between Lie algebras and groups appears also as a problem of great interest in both mathematics and modern theoretical physics.
Examples of these procedures are known as \textit{contractions} \cite{Segal,IW,Saletan} and \textit{deformations} \cite{def1,def2,def3} of Lie algebras, both sharing the property of preserving the dimension of the algebras involved in the process. 

The idea that lead to the concept of contractions was first introduced in Ref. \cite{Segal} and consists in the observation that if two physical theories are related by means of a limit process, then the corresponding symmetry groups under which those theories are invariant should be related through a limit process too. For example, the Newtonian mechanics can be obtained from special relativity by taking $c\rightarrow \infty$ (where $c$ is the speed of light) and thus, the non-relativistic limit that brings the Poincaré algebra to the Galilean algebra is a good example of what is called a contraction process. On the other part, a deformation can be regarded as the inverse of a contraction, which means that in the previous example the Poincaré algebra is a deformation of the Galilean algebra. However, in the present article we will not work with deformations, but rather with
generalizations of the contraction method.

The contractions were formally introduced in Ref. \cite{IW} in a way that nowdays it is known as In\"{o}n\"{u}-Wigner (IW) contraction. The contraction $\mathcal{G}_{c}$ of an algebra $\mathcal{G}$ is made with respect to a subalgebra $\mathcal{L}_{0}$ of $\mathcal{G}$. The procedure bassically consists in rescaling the generators of $\mathcal{G}/\mathcal{L}_{0}$ by a parameter $\lambda$ and then to perform a singular limit for that parameter. As a result, the generators of $\mathcal{G}/\mathcal{L}_{0}$ becomes abelian and the subalgebra acts on them. Further details and explicit examples, including the mentioned relation between Galileo and Poincaré algebras, can be found in Refs. \cite{Barut,Gilmore}. Remarkably, if the original algebra have a more general subspace structure, it is possible to perform more general contractions which are known as Weimar-Woods (WW) contractions \cite{WW}. 

Recently, an interesting generalization has been parallelly introduced in the context of string theory \cite{hs} and supergravity \cite{aipv1,aipv2,aipv3}.
This procedure, known as \textit{expansion} method, is not only able to reproduce the WW contractions when the dimension is preserved in the process, but also may lead to expanded algebras whose dimension is higher than the original one (a result that cannot be obtained by any contraction process).
Another distinguishing feature is that the algebra is described in its dual formalism, i.e., in terms of the Maurer-Cartan (MC) forms on the manifold of its associated Lie group (a good introduction to the dual formulation of Lie algebras can be found in Chapter 5.6 of \cite{Nakahara}).
Instead of rescaling the generators by a real parameter $\lambda$, as it is usually made in the contractions methods, the rescaling is performed on some of the group coordinates. As a consequence, the MC forms can be expanded as power series in $\lambda$ that under certain conditions can be trucated in such a way that assures the closure of a new bigger algebra. 

In this work we deal with an even more general procedure called \textit{S-expansion} \cite{irs} that not only reproduces the results of expansion method described before (which means in turn that also reproduces all WW contractions), but also allows to establish relations between Lie algebras that cannot be obtained by the previous expansion procedure.
Instead of performing a rescaling of some of the algebra generators or the group coordinates, it combines the structure constants of the algebra $\mathcal{G}$ with the
inner multiplication law of an abelian semigroup $S$ in order to define the Lie bracket of a new S-expanded algebra. 
Under certain conditions, it is possible to extract smaller algebras which are called \textit{resonant subalgebras} and \textit{reduced algebras}. On that stage, the S-expansion is able to reproduce the results of the previous expansion method \cite{hs,aipv1}, for a particular family of semigroups denoted by $S_{E}^{\left(  N\right)}$.

An important advantage of the S-expansion method is related with the construction of invariant tensors (and their duals known as Casimir operators) whose full classification is known only\footnote{The classification of all invariant tensors and Casimirs for non-semisimple algebras is still an open problem in Lie theory.} for semisimple Lie algebras. The standard procedure to construct an invariant tensor of range $r$ is to use the symmetrized trace (or supertrace for superalgebras) for the product of $r$ generators in some matrix representation. 
The good feature of the S-expansion is that if we know the invariant tensors of a certain semisimple Lie algebra, then the mechanism gives the invariant tensors for the expanded algebras even if they are not semisimple (the same result was extended in Ref. \cite{Diaz:2012zza} for the Casimir operators). Those invariant tensors of the S-expanded algebra are in general different from the symmetrized trace so it would be interesting to analyze if the S-expansion could help to solve the classification of invariant tensors for non-semisimple Lie algebras. 
Indeed, it has already been supposed in the early 60's that solvable Lie algebras could be obtained as contractions of semisimple algebras of the same dimension \cite{Zaitsev,Celeghini}. The incorrectness of that conjecture was shown in Ref. \cite{Nesterenko} and, recently, that problem has been revisited in terms of S-expansions \cite{Nesterenko2012}. However, to know whether S-expansions are able to fit the classification of non-semisimple Lie algebras and their invariant tensors, remains an open problem.
Independently of the answer, what we do know is that the S-expansion gives invariant tensors that are in general different from the symmetrized trace and this fact has already been useful for the construction of Chern-Simons (CS) gauge theories of (super)gravity \cite{Zanelli:2005sa,Izaurieta:2006aj} and their interrelations.

The dual formulation \cite{irs2} in terms of MC forms has also been very useful, as it allows to perform the S-expansion procedure directly on the Lagrangian of a given gravity theory. This has been used in Refs. \cite{Edelstein:2006se, Izaurieta:2009hz,Concha:2013uhq, Concha:2014vka,Concha:2014zsa} to show that General Relativity (GR) in even and odd dimensions may emerge as a special limit of a Born Infeld \cite{DG} and CS Lagrangian respectively.
Black hole and cosmological solutions has been studied for gravity theories based on expanded algebras Refs. \cite{Quinzacara:2012zz,Quinzacara:2013uua,Crisostomo:2014hia,Crisostomo:2016how}, as well as some aspects about their non relativistic limits \cite{Gonzalez:2016xwo}.
In addition, the S-expansion has also been extended to other mathematical structures, like the case of higher order Lie algebras and infinite dimensional loop algebras \cite{Caroca:2010ax,Caroca:2010kr,Caroca:2011zz}.

At the begining, the applications considered only the specific family of semigroups that allows to reproduce the previous expansion method \cite{hs,aipv1}. These semigroups were denoted by $S_{E}^{(N)}$ and lead to the so called $\mathfrak{B}_{N}$ algebras \cite{irs}. 
The use of others abelian semigroups to perform S-expansions of Lie algebras was first considered in Ref. \cite{Caroca:2011qs}. 
It was shown that some Bianchi algebras \cite{bian} can be obtained as S-expansions form the two dimensional isometries acting transitively in a two-dimensional space. The semigroups that allows to obtain those relations are not all belonging to the $S_{E}^{(N)}$ family and thus, it is clear that these results can only be obtained in the context of the S-expansion, i.e., they cannot be reached by using the previous expansion procedure \cite{aipv1}.
This procedure was then used in Ref. \cite{Diaz:2012zza} to show that the
semisimple version of the so called Maxwell algebra (introduced in \cite{Soroka:2006aj,Durka:2011nf,Durka:2011gm}) 
can be obtained as an expansion of the AdS algebra. Later, this result was generalized in Refs.
\cite{Salgado:2014qqa,Concha:2016hbt} to new families of semigroups generating algebras denoted by
$\mathfrak{C}_{N}$ and $\mathfrak{D}_{N}$ which have been useful to construct new (super)gravity models \cite{Salgado:2014jka,Fierro:2014lka,Concha:2014xfa,Concha:2014tca,Concha:2015tla,Concha:2015woa,Ipinza:2016con,Concha:2016zdb}. Other recent applications can also be found in \cite{Durka:2016eun,Penafiel:2016ufo}.

On the other hand, a general study of the properties $S$-expansion with arbitrary semigroups, in the context of the classification of Lie algebras, was performed in \cite{Andrianopoli:2013ooa}. It was shown that under the S-expansions some properties of the original algebra are always preserved while others do not in general. 
The explicit examples that allowed to check the results in Refs. \cite{Caroca:2011qs} and \cite{Andrianopoli:2013ooa} were obtained with a set computing programs 
that, in this work, we have improved and further developed to give them in the form of a Java library \cite{webJava} allowing to perform S-expansions with arbitrary finite abelian semigroups. We will present this library as a handbook with examples and all the necessary information to use its methods.

This work is organized as follows: In section \ref{Preliminars} we introduce
the basic ingredients we will use. First, we give a brief description of discrete semigroups and the
$S$-expansion method. Then we give a brief review about the existing literature
about the classification of non-isomorphic semigroups and finite semigroup
computer programs. We conclude that section with a general description of the library and notation used.
In sections \ref{generating} and \ref{Library_Sexpansion} we describe the library, which consists of a set of classes containing different methods allowing us to perform S-expansions with any given finite abelian semigroup. 
The reader who is not an expert in Java language and/or probably is more interested in applying the computational tools presented in this paper, might skip Sections \ref{generating} and \ref{Library_Sexpansion} and go directly to Section \ref{other_app}. 
There, we describe some of the 45 programs provided in \cite{webJava} (see the list in Appendix \ref{list_ex}) as examples to use the library. With those instructions the user can easily create new programs to perform his own calculations just by changing the inputs and even without knowing about Java language.
Finally, Section \ref{comments} contains some comments about possible new applications.

\section{Preliminars}

\label{Preliminars}

\subsection{Discrete semigroups}
\label{semigroups}

We consider a set of $n$ elements $ S =  \lbrace \lambda_\alpha, \alpha = 1 , \cdots , n  \rbrace$. We say that $S$ is a semigroup if it is equipped with an associative product
\begin{equation*}
\begin{tabular}
[c]{rrrc}%
$\cdot:$ & $S\times S$ & $\rightarrow$ & $S$\\
& $\left(  \lambda_{\alpha},\lambda_{\beta}\right)  $ & $\rightarrow$ &
$\lambda_{\kappa\left(  \alpha,\beta\right)  }$ \,.%
\end{tabular}
\end{equation*}
Notice that:
\begin{itemize}
\item It does not exist necessarily the identity element $e$ satisfying $ \lambda_\alpha \cdot \,e = \lambda_\alpha \;\forall \alpha = 1, \hdots , n $
\item The elements $ \lambda_\alpha $ do not need to have an inverse.
\item If there exists an element $0_S$ such that $ \lambda_\alpha \cdot 0_S = 0_S \; \forall \alpha $ we will call it a \textit{zero element}
\item $n$ is the \textit{order} of the semigroup
\item If $ \lambda_\alpha \cdot \lambda_\beta = \lambda_\beta \cdot \lambda_\alpha $, the discrete semigroup is said to be {\em commutative} or {\em abelian}
\end{itemize} 
We can give the product by means of a multiplication table, a $ n \times n $ matrix 
\begin{equation}
A=%
\begin{pmatrix}
a_{\alpha\beta}%
\end{pmatrix}
\equiv%
\begin{pmatrix}
\lambda_{\alpha}\cdot\lambda_{\beta}%
\end{pmatrix}
\,,\label{mult_table}%
\end{equation}
with entries in $\lambda_\alpha$.
Thus, a visual way to describe a semigroup is given by,
\begin{equation}%
\begin{tabular}
[c]{c|ccccc}
& $\lambda_{1}$ & $\cdots$ & $\lambda_{\beta}$ & $\cdots$ & $\lambda_{n}%
$\\\hline
$\lambda_{1}$ &  &  &  &  & \\
$\vdots$ &  &  &  &  & \\
$\lambda_{\alpha}$ &  &  & $\lambda_{\kappa\left(  \alpha,\beta\right)  }$ &
& \\
$\vdots$ &  &  &  &  & \\
$\lambda_{n}$ &  &  &  &  &
\end{tabular}
\,.\label{Ex_s_table}%
\end{equation}
For instance, it allows us to check easily if a semigroup is commutative, because in that case its multiplication table is symmetric. 

An informal way (but useful for our purposes) of expressing the multiplication group law is by means of the quantities $ K_{\alpha \beta}^\kappa $, called \textit{selectors}, which are defined in the following way 
\begin{equation}
K_{\alpha \beta}^\kappa = 
\begin{cases}
1 \; \hbox{if} \; \lambda_\alpha \cdot \lambda_\beta = \lambda_\kappa \,, \\
0 \; \hbox{if} \; \lambda_\alpha \cdot \lambda_\beta \neq \lambda_\kappa \,. \\
\end{cases}
\label{sel1}
\end{equation} 
Then, the semigroup law can be expressed as follows:

\begin{equation}
\lambda_\alpha \cdot \lambda_\beta = K_{\alpha \beta}^\kappa \lambda_\kappa \,. \label{sel2}
\end{equation} 
As shown explicitly in Ref. \cite{irs}, from the associativity and closure of the semigroup it follows that the selectors provide a matrix representation
for $S$ and this fact that will be used in the next section to define the S-expansion method.

\paragraph*{Isomorphisms of semigroups}
Consider the semigroups given by the following multiplication tables, 
\begin{equation}%
\begin{tabular}
[c]{c|cc}
& $\lambda_{1}$ & $\lambda_{2}$\\\hline
$\lambda_{1}$ & $\lambda_{1}$ & $\lambda_{1}$\\
$\lambda_{2}$ & $\lambda_{1}$ & $\lambda_{1}$%
\end{tabular}
\,,\ \
\begin{tabular}
[c]{c|cc}
& $\lambda_{1}$ & $\lambda_{2}$\\\hline
$\lambda_{1}$ & $\lambda_{2}$ & $\lambda_{2}$\\
$\lambda_{2}$ & $\lambda_{2}$ & $\lambda_{2}$%
\end{tabular}
\,.\label{1st_ex_isom}%
\end{equation}
These two semigroups have exactly the same structure if we rename $ \lambda_1 $ by $ \lambda_2 $ and viceversa. This is an example of an isomorphism of semigroups. 

The group of isomorphisms between semigroups of order $n$ is isomorphic to the group of permutations of $n$ elements $\Sigma_n$. For simplicity, we choose to represent a permutation by
\begin{equation}
\left(  \lambda_{\alpha_{1}}\,\lambda_{\alpha_{2}}\cdots\lambda_{\alpha_{n}%
}\right)  \label{perm}%
\end{equation}
which means change $\lambda_{1}$ by $\lambda_{\alpha_{1}}$ , change
$\lambda_{2}$ by $\lambda_{\alpha_{2}}$ , $\cdots$ , and finally change
$\lambda_{n}$ by $\lambda_{\alpha_{n}}$.
Then, isomorphisms between semigroups can be defined in terms of their multiplication tables. 
Let $A =
\begin{pmatrix}
a_{\alpha \beta}%
\end{pmatrix}
$ and $B =
\begin{pmatrix}
b_{\alpha \beta}%
\end{pmatrix}
$
be the multiplication tables of two  semigroups of order $n$. According to the definitions given in Ref. \cite{n6-2}, $ A $ and $ B $ describe two isomorphic semigroups if there exists a permutation $ \sigma \in \Sigma_n $ such that
\begin{equation}
A \sigma= B \longleftrightarrow b_{\alpha \beta} = \sigma( a_{\sigma^{-1}(\alpha) ,
\sigma^{-1}(\beta)}) \, \, \, \,  \forall \alpha,\, \beta =1, \hdots,n.
\label{isom}
\end{equation}
If, instead, we have
\begin{equation}
b_{\alpha \beta} = \sigma( a_{ \sigma^{-1}(\beta) , \sigma^{-1}(\alpha)})
\label{antiisom}
\end{equation}
we say that $ A $ and $B $ are related by an {\em anti-isomorphism}.

\subsection{$S$-Expansion of Lie algebras}
\label{sexpp}

Here we briefly describe the general abelian semigroup expansion procedure (S-expansion for short). We refer the interested reader to Refs.\cite{irs} for further details.

First, we need to consider a Lie algebra $\mathcal{G}$ with generators $\{X_{i}\}$ and Lie
bracket
\begin{equation}
\left[  X_{i},X_{j}\right]  =C_{ij}^{\,\,\,\,k}X_{k}\,. \label{p1}%
\end{equation}
where $C_{ij}^{\,\,\,\,k}$ are the structure constants. Next, we need a finite abelian semigroup $S=\left\{  \lambda_{\alpha}\right\}  $, whose multiplication law is given in terms of the selectors $K_{\alpha\beta}^{\gamma}$ defined in Eqs. (\ref{sel1},\ref{sel2}). According to Theorem~3.1 from ref.\cite{irs}, the direct product
\begin{equation}
\mathcal{G}_{S}\text{\ }\mathcal{=}\text{\ }S \otimes\mathcal{G}
\label{z1}%
\end{equation}
is also a Lie algebra, which is called \textit{expanded algebra}. In the proof it can be seen that the commutativity property of the semigroup is crucial for the Jacobi identity to be satisfied in $\mathcal{G}_{S}$.
The elements of this expanded algebra are denoted by
\begin{equation}
X_{\left(  i,\alpha\right)  }=\lambda_{\alpha} \otimes X_{i}\,, 
\label{z2}%
\end{equation}
where $\otimes$ is the Kronecker product of the matrix representations of the generators $X_{i}\in \mathcal{G}$ and the semigroup elements
$\lambda_{a} \in S$. 
The Lie bracket in $\mathcal{G}_{S}$ is defined as
\begin{equation}
\left[  T_{\left(  i,\alpha\right)  },T_{\left(  j,\beta\right)  }\right]
=C_{\left(  i,\alpha\right)  \left(  j,\beta\right)  }^{\left(
k,\gamma\right)  }T_{\left(  k,\gamma\right)  }
\equiv \lambda_{\alpha}\cdot\lambda_{\beta}\otimes\left[  T_{i},T_{j}\right] \label{z3}%
\end{equation}
and therefore the structure constants of the expanded algebra $\mathcal{G}_{S}$ are fully determined by the selectors and the structure constants of the original Lie algebra $\mathcal{G}$, i.e., 
\begin{equation}
C_{\left(  i,\alpha\right)  \left(  j,\beta\right)  }^{\left(  k,\gamma
\right)  }=K_{\alpha\beta}^{\gamma}C_{ij}^{k}\,. %
\end{equation}

There are different cases in which it is possible to systematically extract smaller algebras from $S\otimes\mathcal{G}$. One of them it occurs when the Lie algebra has a decomposition in a direct sum of vectorial subspaces $\mathcal{G}=\bigoplus_{p\in I}V_{p}$, where $I$ is a set of indices encoding the information of the internal subspace structure of the algebra through the mapping\footnote{Here
$2^{I}$ stands for the set of all subsets of $I$.} $i:I\otimes I\rightarrow
2^{I}$ and the relation
\begin{equation}
\left[  V_{p},V_{q}\right]  \subset
{\displaystyle\bigoplus\limits_{r\in i\left(  p,q\right)  }}
V_{r}\,. \label{CDres1}
\end{equation}
If the semigroup $S$ has a decomposition in subsets $S_{p}$, $S=\bigcup_{p\in
I}S_{p}$, satisfying the condition\footnote{Here
$S_{p}\cdot S_{q}$ denotes the set of all the products of all elements from
$S_{p}$ with all elements from $S_{q}$.}
\begin{equation}
S_{p}\cdot S_{q}\subset
{\displaystyle\bigcap\limits_{r\in i\left(  p,q\right)  }}
S_{r} \,,
\label{CDres2}
\end{equation}
which is said to be \textit{resonant} with respect to the subspace structure (\ref{CDres1}) of the algebra, then the subset
\begin{equation}
\mathcal{G}_{S,R}=\bigoplus_{p\in I}S_{p}\otimes V_{p} \label{gen_sub}%
\end{equation}
is a Lie algebra by itself, which is called \textit{resonant subalgebra} of $\mathcal{G}_{S}$ (see Theorem~4.2 from
ref.~\cite{irs}).

Before explaining the next procedures to extract a smaller algebra from
$\mathcal{G}_{S}$ we need to briefly explain what is called a
\textit{reduction} of an algebra. Supose that we have a Lie algebra with a
subspace decomposition $\mathcal{G}=V_{0}\oplus V_{1}$. If the condition
$\left[  V_{0},V_{1}\right]  \subset V_{1}$ is statisfied, then it is
possible to show that the structure constants on $V_{0}$ satisfies the Jacobi
identity by themselves (the proof can be found in Chapter 4.1.2 of Ref.~\cite{irs_IZ}). The
structure constants whose indices take values only on $V_{0}$ defines then a Lie algebra, which is
called reduced algebra $\left\vert V_{0}\right\vert $. Notice that this
definition does not require that $V_{0}$ is a subalgebra.

Now, the second case in which is possible to obtain a smaller algebra occurs
when there is a zero element in the semigroup $S=\left\{  \lambda_{\alpha
},0_{S}\right\}  $. In that case the commutation relations of $\mathcal{G}%
_{S}$ are given by,%
\begin{align}
\left[  X_{\left(  i,\alpha\right)  },X_{\left(  j,\beta\right)  }\right]   &
=C_{ij}^{k}K_{\alpha\beta}^{\gamma}X_{\left(  k,\gamma\right)  }+C_{ij}%
^{k}K_{\alpha\beta}^{0}X_{\left(  k,0\right)  }\,,\nonumber\\
\left[  X_{\left(  i,0\right)  },X_{\left(  j,\beta\right)  }\right]   &
=C_{ij}^{k}X_{\left(  k,0\right)  }\,,\nonumber\\
\left[  X_{\left(  i,0\right)  },X_{\left(  j,0\right)  }\right]   &
=C_{ij}^{k}X_{\left(  k,0\right)  }\,.\label{red_0}%
\end{align}
Thus, the expanded algebra admits a decomposition $\mathcal{G}_{S}=V_{0}\oplus
V_{1}$, with \
\begin{align*}
V_{0}  & =\left(  S/\left\{  0_{S}\right\}  \right)  \otimes\mathcal{G}%
=\left\langle \left\{  X_{\left(  i,\alpha\right)  }\right\}  \right\rangle
\,,\\
V_{1}  & =\left\{  0_{S}\right\}  \otimes\mathcal{G}=\left\langle \left\{
X_{\left(  i,0\right)  }\right\}  \right\rangle \,,
\end{align*}
which clearly satisfies the reduction condition $\left[  V_{0},V_{1}\right]  \subset V_{1}$. According
to the definition ~III.2 of Ref.~\cite{irs}, the structure
constants $C_{\left(  i,\alpha\right)  \left(  j,\beta\right)  }^{\left(
k,\gamma\right)  }=C_{ij}^{k}K_{\alpha\beta}^{\gamma}$ satisfy the Jacobi
identity on $V_{0}$ and then the commutators%
\begin{equation}
\left[  X_{\left(  i,\alpha\right)  },X_{\left(  j,\beta\right)  }\right]
=C_{ij}^{k}K_{\alpha\beta}^{\gamma}X_{\left(  k,\gamma\right)  }%
\,,\label{red_1}%
\end{equation}
define by themselves a Lie algebra, which is called the $0_{S}$%
-\textit{reduced algebra} $\mathcal{G}_{S,\text{red}}$.
Thus, the reduction process is then equivalent to remove the whole
$0_{S}\otimes\mathcal{G}$ sector from the expanded algebra. As we can see from
Eqs. (\ref{red_0}), the reduced algebra defined in this way is not a
subalgebra of $\mathcal{G}_{S}$.

The existence of a resonant decomposition (a resonance for short) and a zero element for the semigroup are mutually independent issues and, consequently, the same is true for the extraction of a resonant subalgebra and a reduced algebra. This is, there are semigroups having resonances but with no zero element and vice versa. Thus, the third way to obtain smaller algebras happens when a given semigroup has simultaneously both properties. In that case it is possible to perform a reduction of the resonant subalgebra and the resulting algebra, denoted by $\mathcal{G}_{S,R,\text{red}}$, is called a $0_{S}$\textit{-reduced resonant subalgebra}.

A forth way to extract smaller algebras from $\mathcal{G}_{S}$ is called
\textit{resonant reduction}. It can be applied when the semigroup have a
resonant decomposition $S=\cup_{p\in I}S_{p}$, satisfying Eq. (\ref{red_1}) for which, in addition, each subset $S_{p}$ admits a partition
$S_{p}=\hat{S}_{p}\cup\check{S}_{p}$ such that
\begin{equation}
\check{S}_{p}\cap\hat{S}_{q}=\phi \,, \label{rr1}%
\end{equation}%
\begin{equation}
\check{S}_{p}\times\hat{S}_{q}\subset\cap_{r\in i_{\left(  p,q\right)  }}%
\hat{S}_{r} \,. \label{rr2}%
\end{equation}
This partition induce a descomposition $\mathcal{G}_{S,R}=\mathcal{\check{G}%
}_{S,R}\oplus\mathcal{\hat{G}}_{S,R}$ on the resonant subalgebra where,
\begin{equation}
\mathcal{\check{G}}_{S,R}=
{\displaystyle\bigoplus\limits_{p\in I}}
\check{S}_{p}\otimes V_{p} \,,\label{rr3}%
\end{equation}%
\begin{equation}
\mathcal{\hat{G}}_{S,R}=
{\displaystyle\bigoplus\limits_{p\in I}}
\hat{S}_{p}\otimes V_{p} \,.\label{rr4}%
\end{equation}
Using conditions (\ref{rr1}) and (\ref{rr2}) it can be shown that $\left[
\mathcal{\check{G}}_{S,R},\mathcal{\hat{G}}_{S,R}\right]  \subset
\mathcal{\hat{G}}_{S,R}$. This implies that $\left\vert \mathcal{\check{G}%
}_{S,R}\right\vert $\ is a reduction of the resonant subalgebra $\mathcal{G}%
_{S,R}$ which is called the \textit{resonant reduction}.

Interestingly, the $0_{S}$-reduction of a resonant subalgebra can be regarded
as a particular case of the resonant reduction. Indeed, consider a semigroup
$S$ with a $0_{S}$ element and a resonant decomposition $S=\cup_{p\in I}S_{p}$
such that $0_{S}\in S_{p}$ for each $p\in I$. If $\mathcal{G}_{S,R}%
=\oplus_{p\in I}S_{p}\otimes V_{p}$ is the corresponding resonant subalgebra,
then the following partition
\[
S_{p}=\hat{S}_{p}\cup\check{S}_{p}\text{ \ with }\ \hat{S}_{p}=\left\{
0_{S}\right\}  \text{ \ and \ }\check{S}_{p}=S_{p}-\left\{  0_{S}\right\} \,,
\]
satisfies the conditions (\ref{rr1}) y (\ref{rr2}) and therefore, in this
particular case, the resonant reduction $\left\vert \mathcal{\check{G}}%
_{S,R}\right\vert $ of $\mathcal{G}_{S,R}$ coincides with the $0_{S}$-reduction of $\mathcal{G}_{S,R}$. Further details and explicit examples can
be found in Ref. \cite{irs}. 

A last independent way to to extract smaller algebras from $\mathcal{G}_{S}$
\ can be done with the so called \textit{H-condition} introduced in Ref.
\cite{Gonzalez:2014tta} that applies when the semigroup is a cyclic group of
even order. However, as that procedure applies only for that family of
semigroups, the thechniques developed in this work are not necessarily needed
for that case.

\subsection{$S$-Expansions and the classification of Lie algebras}
\label{gen_prop}
\begin{figure}[th]
\centering
\includegraphics[scale=0.40]{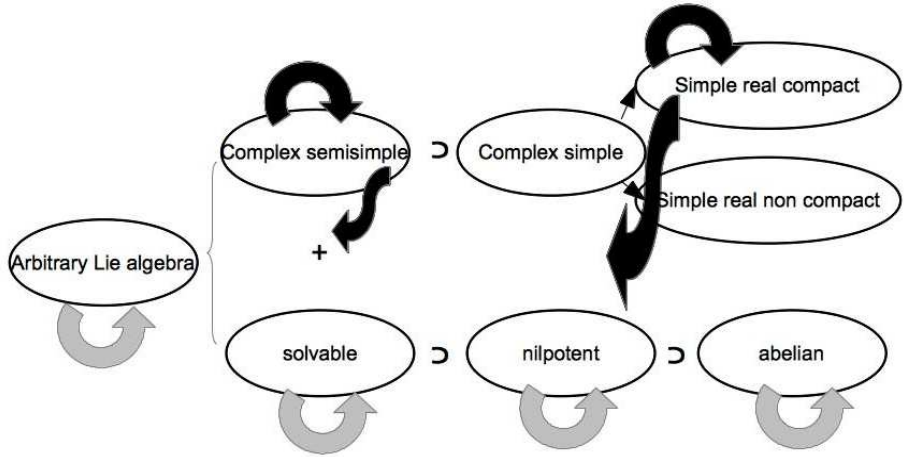}
\caption{Action of the expansion methods on the scheme of the Theory of the Classification of Lie
Algebras.}%
\label{fig:fig1}%
\end{figure}

The properties of the S-expansion in the general context of the classification of Lie algebras has been studied in Ref. \cite{Andrianopoli:2013ooa}. It was shown that abelian, solvable and nilpotent algebras under expansions with any semigroup remains to be respectively abelian, solvable and nilpotent (see figure \ref{fig:fig1}). However, for semisimple and compact algebras the situation is different. It was shown that the quantity
\begin{equation}
\mathbf{g}^{S}\equiv\left(  g_{\alpha\beta}^{S}\right)  =\left(  K_{\alpha\gamma}^{\lambda}K_{\beta\lambda}^{\gamma}\right)\,,  \label{gs}%
\end{equation}
called semigroup metric, can be used to predict if properties like semisimplicity and compactness are preserved or broken under the S-expansion. To see this, let us first remind that:
\begin{itemize}
	\item A Lie algebra $\mathcal{G}$ is \textit{semisimple} if and only if the Killing-Cartan (KC) metric,
	\begin{equation} 
\mathbf{g}\equiv\left(  g_{ij}\right)  =\left(  C_{ik}^{l}C_{jl}^{k}\right)\,, \label{KC_def}%
\end{equation}
	is non degenerate, i.e., if $\det\left(  g_{ij}\right)  \neq0$.
	\item The KC metric is diagonalizable so if we denote by $\left(  \mu_{i}\right)  $ its spectra of eigenvalues, then a semisimple Lie algebra $\mathcal{G}$ is \textit{compact} if and only if $\mu_{i}<0$.
\end{itemize}

Now, the Killing-Cartan (KC) metric of an S-expanded Lie algebra is given by
\begin{equation}
g_{\left(  i,\alpha\right)  \left(  j,\beta\right)  }=C_{\left(
i,\alpha\right)  \left(  k,\gamma\right)  }^{\left(  l,\lambda\right)
}C_{\left(  j,\beta\right)  \left(  l,\lambda\right)  }^{\left(
k,\gamma\right)  }=K_{\alpha\gamma}^{\lambda}K_{\beta\lambda}^{\gamma}%
C_{ik}^{l}C_{jl}^{k}=g_{\alpha\beta}^{S}g_{ij}\,.\label{mario_0}%
\end{equation}
This means that $\mathbf{g}^{E}$ is the Kronecker product of $\mathbf{g}^{S}$
and $\mathbf{g}$,
\begin{equation}
\mathbf{g}^{E}\equiv\left(  g_{\left(  i,\alpha\right)  \left(  j,\beta
\right)  }\right)  =\mathbf{g}^{S}\otimes\mathbf{g}\,,\label{KCexp}%
\end{equation}
with the last two matrices being diagonalizable. Denoting respectively by $\left(
\xi_{\alpha}\right)  $ and $\left(  \mu_{i}\right)  $ the spectra of
eigenvalues of $\mathbf{g}^{S}$ and $\mathbf{g}$ we know, from the general
theory of Kronecker products, that:

\begin{itemize}
\item the eigenvalues of $\mathbf{g}^{E}$\ are $\left(  \xi_{\alpha}\mu
_{i}\right)  $;

\item $\det\left(  \mathbf{g}^{E}\right)  =\det\left(  \mathbf{g}^{S}\right)
^{\dim\left(  \mathcal{G}\right)  }\det\left(  \mathbf{g}\right)  ^{n}$ where
$n$ is the order of the semigroup.
\end{itemize}

Therefore, $\mathcal{G}$ is semisimple ($\det\left(  \mathbf{g}\right)  \neq0$),
$\mathcal{G}_{S}$ is semisimple if and only if $\det\left(  \mathbf{g}%
^{S}\right)  \neq0$. In addition, if $\mathcal{G}$ is compact ($\mu_{i}<0$) then $\mathcal{G}%
_{S}$ is compact only if $\xi_{\alpha}>0$. In other words, the real form of
$\mathcal{G}_{S}$ strongly depends on the signs of $\xi_{\alpha}$.

A similar analysis and the definition of the KC metric for the resonant subalgebra and reduced algebra that will be used in our library (see Section \ref{Library_Sexpansion}), can be found in Section 3.2 of Ref.  \cite{Andrianopoli:2013ooa}. 
In terms of the eigenvalues of the matrices $\mathbf{g}^{S}$, $\mathbf{g}^{S}\left(  S_{p}\right)  $,
and $\mathbf{g}^{\text{red}}\left(S_{p}\right)  $ it was shown that under the action of the $S$-expansion procedure some
properties of the Lie algebras (like commutativity, solvability and
nilpotency) are always preserved while others (like semi-simplicity and compactness)
are not in general. 
This was analyzed for: the expanded algebra $\mathcal{G}_{S}$, the resonant subalgebra $\mathcal{G}_{S,R}$ and the reduced algebras $\mathcal{G}_{S,\text{red}}$ and $\mathcal{G}_{S,R,\text{red}}$.
The same analysis was done for the expansion of a general Lie algebra $\mathcal{G}$ on its Levi-Malcev decomposition $\mathcal{G=}\mathcal{N}\oplus_{s}\mathcal{S}$, where $\oplus_{s}$ represents the semidirect sum of the semisimple subalgebra $\mathcal{S}$ and the maximal solvable ideal $\mathcal{N}$ (also known as radical). The results can be summarized as follow:
\begin{align*}
&  \text{\textbf{Properties preserved by the action of the} }%
S\text{\textbf{-expansion process}}\\
&
\begin{tabular}
[c]{|l|l|l|l|}\hline
\textbf{Original }$\mathcal{G}$ & \textbf{Expanded}$\mathcal{G}_{S}$ &
\textbf{Resonant }$\mathcal{G}_{S,R}$ & \textbf{Reduced }$\mathcal{G}%
_{S,R}^{\text{red}}$\\\hline
abelian & abelian & abelian & abelian\\\hline
solvable & solvable & solvable & solvable\\\hline
nilpotent & nilpotent & nilpotent & nilpotent\\\hline
compact & $\text{arbitrary}$ & $\text{arbitrary}$ & $\text{arbitrary}$\\\hline
$%
\begin{array}
[c]{c}%
\text{semisimple}\\
\mathcal{G=S}%
\end{array}
$ & $%
\begin{array}
[c]{c}%
\text{arbitrary}\\
\mathcal{G}_{S}\mathcal{=}\mathcal{N}_{\text{exp}}\oplus_{s}\mathcal{S}_{\text{exp}}%
\end{array}
$ & $%
\begin{array}
[c]{c}%
\text{arbitrary}\\
\mathcal{G}_{S,R}\mathcal{=}\mathcal{N}_{\text{exp,}R}\oplus_{s}\mathcal{S}%
_{\text{exp,}R}%
\end{array}
$ & $%
\begin{array}
[c]{c}%
\text{arbitrary}\\
\mathcal{G}_{S,R}^{\text{red}}\mathcal{=}\mathcal{N}_{\text{exp,}R}^{\text{red}}%
\oplus_{s}\mathcal{S}_{\text{exp,}R}^{\text{red}}%
\end{array}
$\\\hline
$%
\begin{array}
[c]{c}%
\text{arbitrary}\\
\mathcal{G=}\mathcal{N}\oplus_{s}\mathcal{S}%
\end{array}
$ & $%
\begin{array}
[c]{c}%
\text{arbitrary}\\
\mathcal{G}_{S}\mathcal{=}\mathcal{N}_{\text{exp}}\oplus_{s}\mathcal{S}_{\text{exp}}%
\end{array}
$ & $%
\begin{array}
[c]{c}%
\text{arbitrary}\\
\mathcal{G}_{S,R}\mathcal{=}\mathcal{N}_{\text{exp,}R}\oplus_{s}\mathcal{S}%
_{\text{exp,}R}%
\end{array}
$ & $%
\begin{array}
[c]{c}%
\text{arbitrary}\\
\mathcal{G}_{S,R}^{\text{red}}\mathcal{=}\mathcal{N}_{\text{exp,}R}^{\text{red}}%
\oplus_{s}\mathcal{S}_{\text{exp,}R}^{\text{red}}%
\end{array}
$\\\hline
\end{tabular}
\end{align*}

Figure \ref{fig:fig1} illustrates the general scheme of the classification
theory of Lie algebras and the arrows represent the action of the expansion
method on this classification. \textit{Grey arrows} are used when the
expansion method maps algebras of one set on to the same set, i.e., they
preserve some specific property. On the other hand, \textit{black arrows}
indicate that the expansion methods can map algebras of one specific set on to
the same set and also can lead us outside the set, to an algebra that will
have a Levi-Malcev decomposition.

The Cartan decomposition of the expanded algebra was also obtained when the
expansion preserves compactness. To check all these theoretical results, an
example was given by studying all the possible expansions of the semisimple
algebra $\mathfrak{sl}(2,\mathbb{R})$ with semigroups of order up to $6$. 
This checking was made with some computer programs that has been the starting point to construct the Java library that we will present in the next Sections. 
Before showing how the library is constructed and how it works, we give a brief review of the classification of semigroups.

\subsection{Review of finite semigroups programs}
\label{hist_semig}

Discrete semigroups have been a subject of intensive research in Mathematics and its
classification have been made by many different authors (see e.g. \cite{n6-2,n4,n5,n6-1,n6-3,n7,n8,n9-1,n9-2,n9-3,n9-4} and references therein).
In particular, the number of finite non-isomorphic semigroups of order $n$ is given in the following table:

\begin{equation}%
\begin{tabular}
[c]{|l|l|l}\cline{1-2}%
order & $Q=\ $\# semigroups & \\\cline{1-2}%
1 & 1 & \\\cline{1-2}%
2 & 4 & \\\cline{1-2}%
3 & 18 & \\\cline{1-2}%
4 & 126 & [Forsythe '54]\\\cline{1-2}%
5 & 1,160 & [Motzkin, Selfridge '55]\\\cline{1-2}%
6 & 15,973 & [Plemmons '66]\\\cline{1-2}%
7 & 836,021 & [Jurgensen, Wick '76]\\\cline{1-2}%
8 & 1,843,120,128 & [Satoh, Yama, Tokizawa '94]\\\cline{1-2}%
9 & 52,989,400,714,478 & [Distler, Kelsey, Mitchell '09]\\\cline{1-2}%
10 & \textbf{12,418,001,077,381,302,684} & [Distler, Jefferson, Kelsey,
Kotthoff '16]\\\cline{1-2}%
\end{tabular}
\ \label{hist}%
\end{equation}

As shown in the table the problem of enumerating the all non-isomorphic
finite semigroups of a certain order is a non-trivial problem. In fact, the
number $Q$ of semigroups increases very quickly with the order of the semigroup.
Since the original algorithms proposed by Plemmons \cite{n6-1,n6-2,n6-3} to computationally generate non isomorphic semigroups, a lot of work 
has been aided by computers. Remarkably, the order 9 has only
been reached with the aid of supercomputers and the number (but not the explicit list of semigroup tables) for the order 10 has been obtained very recently \cite{S10}.

Particularly, in Ref. \cite{Hildebrant} there were constructed very useful algorithms to make calculations with finite semigroups. 
First, the program \textit{gen.f} generates in lexicographical ordering the lists of all the non-isomorphic semigroups of order $n=1,2,...,8$. 
This means that if we find a semigroup table $\tilde{S}$ which is not contained in one of those lists, then: 
\begin{itemize}
	\item $\tilde{S}$ does not have lexicographical ordering and
	\item $\tilde{S}$ is isomorphic to one and only one table of the lists generated by \textit{gen.f}.
\end{itemize}
In each list, a semigroup $S_{\left(n\right)}^{a}$ of order $n$ is univocally identified by the number $a=1,...,Q$ and the semigroup elements are denoted by $\lambda_{\alpha}$ with $\alpha=1,...,n$.

The second program of Ref. \cite{Hildebrant} is \textit{com.f}, which takes one of these lists generated by \textit{gen.f} and selects only the abelian semigroups. For
example, for $n=2$ the elements are labeled by $\left\{  \lambda_{1}%
,\lambda_{2}\right\}  $ and the program \textit{com.f} gives the following
list of semigroups:%

\begin{equation}%
\begin{tabular}
[c]{l|ll}%
$S_{\left(  2\right)  }^{1}$ & $\lambda_{1}$ & $\lambda_{2}$\\\hline
$\lambda_{1}$ & $\lambda_{1}$ & $\lambda_{1}$\\
$\lambda_{2}$ & $\lambda_{1}$ & $\lambda_{1}$%
\end{tabular}
\ \ \ \text{, }%
\begin{tabular}
[c]{l|ll}%
$S_{\left(  2\right)  }^{2}$ & $\lambda_{1}$ & $\lambda_{2}$\\\hline
$\lambda_{1}$ & $\lambda_{1}$ & $\lambda_{1}$\\
$\lambda_{2}$ & $\lambda_{1}$ & $\lambda_{2}$%
\end{tabular}
\ \ \ \text{, }%
\begin{tabular}
[c]{l|ll}%
$S_{\left(  2\right)  }^{4}$ & $\lambda_{1}$ & $\lambda_{2}$\\\hline
$\lambda_{1}$ & $\lambda_{1}$ & $\lambda_{2}$\\
$\lambda_{2}$ & $\lambda_{2}$ & $\lambda_{1}$%
\end{tabular}
\ \ \label{list1}%
\end{equation}
Note that the semigroup $S_{\left( 2\right)  }^{3}$ is not given in the list
(\ref{list1}) because it is not abelian. 
In the library that will described in the next section, it will be easier to use the full lists of semigroups generated by \textit{gen.f}. When needed,
we will select the abelian ones using a method similar to \textit{com.f}.

\subsection{Description of the library and notation}

\label{descriptionLib}

As a preliminar step, we have used the program \textit{gen.f} of Ref. \cite{Hildebrant} to generate the files \textit{sem.2}, \textit{sem.3}, \textit{sem.4}, \textit{sem.5} and \textit{sem.6} which contain all the non isomorphic semigroups up to\footnote{In principle, with the program \textit{gen.f} it is possible to generate the full lists of non isomorphic semigroups up to order 8. In our case we were able to compute the lists up to order 6. However, various methods of our library are able to perform calculations with semigroups of order higher than 6 like checking associativity, finding zero elements, isomorphisms and resonances (see e.g., Sections \ref{asso_com} and \ref{gen_checkings2}).} order 6. 
As we will see, those files are the input data for many of our programs.

To use the Java library that we present in this work it is necessary to download the linear algebra package \textit{jama.jar}\footnote{In particular, we use the methods belonging to the class \textit{Matrix} of the library \textit{jama.jar}.} from \cite{jama} and the following files from \cite{webJava}: 

\begin{equation}%
\begin{tabular}
[c]{|l|l|}\hline
\textbf{File} & \textbf{Brief description}\\\hline
\textit{data.zip} & Contains the the files \textit{sem.n}\\\hline
\textit{sexpansion.jar} & Is the library itself\\\hline
\textit{examples.zip} & Example programs listed in Appendix \ref{list_ex} and
explained in Section \ref{other_app}\\\hline
\textit{Output\_examples.zip} & Output samples of the example programs\\\hline
\end{tabular}
\ \label{table_files}%
\end{equation}

A \textit{ReadMe.pdf} file is also provided with detailed installation instructions, should they be needed.
\bigskip

Our library is composed of the following \textit{classes} (to see the source code unzip the file \textit{sexpansion.jar}):
\begin{small}
\begin{enumerate}
   \item Semigroup.java
   \item SetS.java
   \item Selector.java
   \item SelectorReduced.java
   \item SelectorResonant.java
   \item SelectorResonantReduced.java
   \item StructureConstantSet.java
   \item StructureConstantSetExpanded.java
   \item StructureConstantSetExpandedReduced.java
   \item StructureConstantSetExpandedResonant.java
   \item StructureConstantSetExpandedResonantReduced.java
\end{enumerate}
\end{small}

Each \textit{class} contains different \textit{methods}. A brief description can be found in the code at the beginning of each class and method. As the full documentation of the library is available in \cite{wiki}, in this article we will describe only the most important methods of these classes (see Sections \ref{generating}-\ref{Library_Sexpansion}).

In addition, there are 45 programs included as examples in the file \textit{examples.zip} whose aim is to show how the methods of our library works. They are named with the prefixes \textit{I}, \textit{II}, \textit{III} which stand for the following classification:

\bigskip

\textbf{\textit{I} \ \ \ :} \ General computations with semigroups,

\textbf{\textit{II} \ \ :} \ Examples of S-expansions of Lie algebras

\textbf{\textit{III} \ :} \ Programs related with the calculations made in Ref. \cite{Andrianopoli:2013ooa}.
\bigskip

The list of these example programs is given in the Appendix \ref{list_ex} and most of them will be described in the section \ref{other_app}.

\paragraph*{The type of resonances implemented on this first version}

The methods that will be described in Sections \ref{generating} and \ref{Library_Sexpansion}, are able to work with algebras having a decomposition $\mathcal{G}=V_{0}\oplus V_{1}$ with the subspace structure given by
\begin{align}
\left[  V_{0},V_{0}\right]   &  \subset V_{0} \,, \nonumber\\
\left[  V_{0},V_{1}\right]   &  \subset V_{1} \,, \nonumber\\
\left[  V_{1},V_{1}\right]   &  \subset V_{0} \,. \label{subspace_structure_}%
\end{align}
Correspondingly, the type of resonant decomposition that will be considered have the form $S=S_{0}\cup S_{1}$ and satisfy
\begin{align}
S_{0}\times S_{0}  &  \subset S_{0} \,, \nonumber\\
S_{0}\times S_{1}  &  \subset S_{1} \,, \nonumber\\ 
S_{1}\times S_{1}  &  \subset S_{0} \,. \label{resonant_cond}%
\end{align}
Then, the resonant subalgebras that will be obtained have the form,
\begin{equation}
\mathcal{G}_{S,R}=\left(  S_{0}\otimes V_{0}\right)  \oplus\left(
S_{1}\otimes V_{1}\right)\,. \label{res_subalg}%
\end{equation}

We will also deal with the case in which the semigroup may have a $0_{S}$ element. Therefore, the library will allow us to extract from $\mathcal{G}_{S}$ the following three types of smaller algebras:
\begin{itemize}
	\item Resonant subalgebras, when the original subalgebra has the subspace decomposition described in (\ref{subspace_structure_}) and the semigroup posseses a resonant decomposition satisfying (\ref{resonant_cond}),
	\item $0_{S}$-Reduced algebras, when the semigroup has a $0_{S}$ element,
	\item $0_{S}$-Reduction of the resonant subalgebra when the semigroup has simultaneously a $0_{S}$ element and the resonant decomposition given by Eq. (\ref{resonant_cond}).
\end{itemize}

In a first stage, the situation described by Eqs. (\ref{subspace_structure_}-\ref{res_subalg}) (particular cases of Eqs. (\ref{CDres1}-\ref{gen_sub})), will be enough for the applications we want to consider in this work. 
However, as our library has an open GNU licence\footnote{This type of licence basically means that anyone can download, use and modify the library. We do appreciate the corresponding citation to our original version.}, we expect that it can be extended in order to perform S-expansions of algebras having more general subspace structures.

\paragraph*{Conventions and notation}

For programming language reasons, the semigroup elements $\lambda_{\alpha}$ must be labeled by their sub-index number,
\begin{equation}
\lambda_{\alpha}\longleftrightarrow\alpha \,,\label{convent}%
\end{equation}%
with $\ \alpha=1,\ldots,n$. With this convention, the clousure is given by
\begin{equation}
\lambda_{\kappa\left(  \alpha,\beta\right)  }=\lambda_{\alpha}\cdot
\lambda_{\beta}\longleftrightarrow\kappa\left(  \alpha,\beta\right)
=\alpha\cdot\beta\,\label{convent1}%
\end{equation}
so that the multiplication table, Eq. (\ref{mult_table}), can be written as
\begin{equation}
A=%
\begin{pmatrix}
a_{\alpha\beta}%
\end{pmatrix}
\equiv%
\begin{pmatrix}
\alpha\cdot\beta
\end{pmatrix}
\,.\label{convent2}%
\end{equation}

Thus, properties like associative and commutativity respectively read,
\begin{equation}
\left(  \alpha\cdot\beta\right)  \cdot\gamma=\alpha\cdot\left(  \beta
\cdot\gamma\right)  \ ,\ \ \alpha\cdot\beta=\beta\cdot\alpha \,, \label{asso_def1}%
\end{equation}
which in terms of the multiplication table, can be expressed as
\begin{equation}
a_{a_{\alpha,\beta},\gamma}=a_{\alpha,a_{\beta,\gamma}}\ ,\ \ a_{\alpha\beta
}=a_{\beta\alpha}\,.\label{asso_def2}%
\end{equation}
An isomorphism, like the one described in Eq. (\ref{perm}), will be then denoted by%
\begin{equation}
\left(  \alpha_{1}\alpha_{2}\cdots\alpha_{n}\right)  \label{perm_not}%
\end{equation}
and means: change $1$ by $\alpha_{1}$, change $2$ by $\alpha_{2}$,
\ldots\ and finally change $n$ by $\alpha_{n}$.

In the codes, latin characteres are used to label both semigroup elements and generators of the Lie algebra and its meaning is specified.

\section{Methods for semigroups calculations}
\label{generating}

\subsection{Loading all the non-isomorphic semigroups}
\label{loading}

Using the conventions and notations adopted in Section \ref{descriptionLib}, we define the \textit{Semigroup} class to represent a discrete semigroup and all the
operations which can be performed with them.

\begin{small}
\begin{lstlisting}[language=Java]
public class Semigroup {
 int[][] data;
 int order;  
 int ID;  
\end{lstlisting}
\end{small}
Obviously the code is bigger\footnote{We remind to the reader that the library, containing the full code of all the programs described here, is available in \cite{webJava}.}, but for space reasons we restrict ourselves only to describe the main parts of the methods and classes.
Thus, we use 3 variables to save the information of a semigroup $S_{\left(  n\right)
}^{a}$: an integer \textit{ID} which correspond to the identifier $a$ of the
semigroup, a second integer \textit{order} which tells us the order $n$ of the
semigroup and a matrix of integers \textit{data} where we save the
multiplication table of the semigroup
$A = 
\begin{pmatrix}
a_{\alpha \beta}%
\end{pmatrix}
\equiv%
\begin{pmatrix}
\alpha \cdot \beta%
\end{pmatrix}$. 

As explained in Section \ref{descriptionLib}, in many of the calculations that will be done is necessary to load the semigroups generated
by the program \textit{gen.f} of Ref. \cite{Hildebrant}. This is done by the method \textit{loadFile} which loads all the semigroups of a given order (up to 
order 6) returning us an array of \textit{Semigroup} objects. 
To load all the avaliable semigroups from order 2 to 6 we use the method \textit{loadFromFile}, which uses the method \textit{loadFile}. 
In Section \ref{other_app} we will show how the method \textit{loadFromFile} is used in different problems.

\subsection{Associativity, commutativity and $0_{S}$-element}
\label{basics}

To check if a given a multiplication table of a set of elements $S=\left\{  {\alpha}\right\}  _{\alpha=1}^{n}$ is a semigroup, we must check if it is associative. This means to check the relations $\left(  \alpha\cdot\beta\right)  \cdot\gamma=\alpha\cdot\left(  \beta
\cdot\gamma\right) $ for all $\alpha,\beta,\gamma=1,\ldots,n$. We perform this task with the method \textit{isAssociative}.

\begin{small}
\begin{lstlisting}[language=Java]
public boolean isAssociative() {
 int i , j , k ;
 for ( i = 0 ; i < order ; ++ i) {
  for ( j = 0 ; j < order ; ++j) {
   for ( k = 0 ; k < order ; ++ k) {
    if( ! (data[i][data[j][k]-1] == data[data[i][j]-1][k])) {
   return false ;
 }}}}
return true;
}
\end{lstlisting}
\end{small}
Basically, this method returns true if a given multiplication table is associative.

On the other hand, as described in section \ref{sexpp}, to perform a S-expansion the semigroup used must be commutative. The method
\textit{isCommutative} returns true if a given semigroup is commutative, i.e., $\alpha \cdot \beta = \beta \cdot \alpha$.

\begin{small}
\begin{lstlisting}[language=Java]
public boolean isCommutative(){
 int i, j;
 for ( i = 0 ; i < order ; ++i) {
  for( j = 0 ; j < order ; ++j ) {
   if( ! (data[i][j] == data[j][i])) {
  return false;
 }}}
 return true;
}
\end{lstlisting}
\end{small}

In section \ref{sexpp} we have explained that when a commutative semigroup has a zero element satisfying $\alpha\cdot0_{S} = 0_{S} \; \forall\alpha$ it is possible to perform a $0_{S}$-reduction of the expanded algebra. 
To be able to automatize that procedure we need a method which can look for the
zero element for any semigroup. The method \textit{findZero} gives the zero element 
when the semigroup does have it, otherwise it returns $-1$ as a result.
\begin{small}
\begin{lstlisting}[language=Java]
public int findZero() {
 int i, j ;
 boolean isZero = false ;
 for ( i = 0 ; i < order ; ++ i ) {
  if ( isZero == true ) {
   return i ;
  }
  j = 0 ;
  isZero = true;
  while ( isZero && (j < order )) {
   if ( data[j][i] != i+1) {
    isZero = false ;
   }
   ++j;
 }}   
 if ( isZero == true ) {
 return order ;
 }
 return -1 ;
}
\end{lstlisting}
\end{small}

On the other hand, it is also useful to have in some cases an auxiliar method to check for the equality of two
given semigroups. The method \textit{isEqualTo} returns true if they are equal.
\begin{small}
\begin{lstlisting}[language=Java]
public boolean isEqualTo ( Semigroup B ) {
 int i,j;
 if ( this.order != B.order ) {
 	return false;
 }
 for ( i = 0 ; i < this.order ; ++i ) {
 	for ( j = 0 ; j < B.order ; ++j) {
   if ( this.data[i][j] != B.data[i][j]){
   	return false;
 }}}
 return true;
}
\end{lstlisting}
\end{small}

In Section \ref{other_app} we will show how the methods \textit{loadFromFile}, \textit{isAssociative}, \textit{isCommutative}, \textit{findZero} and \textit{isEqualTo} are used in different problems.

\subsection{Creating sets and the permutation group $\Sigma_{n} $}
\label{isomorphisms}

As explained in section \ref{semigroups}, the group of isomorphism of the semigroups or order $n$ is
the group of permutations of $n$ elements, $\Sigma_{n}$.
Thus, we define the \textit{SetS} class to represent a permutation.

\begin{small}
\begin{lstlisting}[language=Java]
public class SetS {
 int [] list ;
 int nElements ;
\end{lstlisting}
\end{small}

A \textit{SetS} object contains an integer \textit{nElements} representing the number of elements to which we want to apply the permutation and a \textit{list} where we save the permutation with the notation given in Eq. (\ref{perm_not}) in Section \ref{descriptionLib}. This objects will also serve us to save any set of non
repeated integers, like the one we will use for the resonant decomposition of a discrete semigroup.

Again, for space reasons, we are not able to reproduce the complete definition of the \textit{SetS} class here and only its main methods will be described in what follows (remind that the full library is available on \cite{webJava}).

To create a \textit{SetS} object just from an array of integers we can use the code
\begin{small}
\begin{lstlisting}[language=Java]
public SetS( int [] elements ) {
 list = elements ;
 nElements = elements.length;
}
\end{lstlisting}
\end{small}
while to create a \textit{SetS} of $n$ elements containing the identity permutation in
$\Sigma_{n} $ we should use

\begin{small}
\begin{lstlisting}[language=Java]
public SetS( int n ){
 nElements = n;
 list = new int[n];
 int i ;
 for ( i = 0 ; i < n ; ++i) {
 	list[i] = i +1 ;
}}
\end{lstlisting}
\end{small}

Now, to check if two given semigroups are isomorphic we have to try all the
existent isormorphisms for a given order. This means to use all the elements in
$\Sigma_{n} $. This is performed by the method \textit{allPermutations}, which returns
an array of \textit{SetS} objects containing all the elements in $\Sigma_{n} $.
\begin{small}
\begin{lstlisting}[language=Java]
SetS [] allPermutations( ) {
 SetS result = new SetS(0) ;
 return this.permutationsAux( this , result);
}
\end{lstlisting}
\end{small}
This method uses the auxiliar method \textit{permutationsAux}, which actually performs
most of the work. This is a recursive method which takes an original \textit{SetS} object and reorders its elements in all the possible ways. This is, given
the identity permutation
\[
\left(  1 \, 2 \, \cdots\, n \right)
\]
it returns all the possible ways to permute it, i.e., all the elements of the permutation group $\Sigma_{n}$. 
As we can see in what follows, \textit{permutationsAux} uses the methods \textit{addElement} and \textit{eraseElement} whose definition is obvious and we do not reproduce here.
\begin{small}
\begin{lstlisting}[language=Java]
SetS [] permutationsAux( SetS original, SetS result ) {
int i , j;
SetS original2 , result2 ;
SetS [] list = null;
SetS [] totalList = null;
SetS [] previousList = null;
int N = 0 ;
if ( original.nElements == 0 ) {
 list = new SetS[1];
 list[0] = result ;
 return list ;
 }
 for ( i = 0 ; i < original.nElements ; i++ ){
  result2 = result.addElement( original.elementAt(i) ) ;
  original2 = original.eraseElement(i) ;
  list = permutationsAux( original2 , result2) ;
  N = N + list.length ;
  previousList = totalList ;
  totalList = new SetS[N];
  for ( j = 0 ; j < N - list.length ; ++j) {
   totalList[j] = previousList[j];
  }
  for ( j = 0 ; j < list.length ; ++j) {
   totalList[ j + N - list.length ] = list[j];
 }}
 return totalList;
}
\end{lstlisting}
\end{small}

As explained in section \ref{semigroups}, two given semigroups are isomorphic or anti-isomorphic if there exist a permutation $ \sigma \in \Sigma_n $ that relates them through the Eqs. (\ref{isom}) and (\ref{antiisom}).
Here we will describe two basic operations related with isomorphisms. First, the method which applies a given isormorphism to a semigroup is \textit{permuteWith}:
\begin{small}
\begin{lstlisting}[language=Java]
public Semigroup permuteWith( SetS s ) {
int i,j;
int [][] matrix = new int[this.order][ this.order];
SetS inverse  = s.inversePermutation() ;
for ( i = 0 ; i < this.order ; ++i) {
 for ( j = 0 ; j < this.order ; ++j) {
  matrix[i][j] = this.data[ inverse.elementAt(i) - 1] 
   [ inverse.elementAt(j) -1] ;
 }}
 for ( i = 0 ; i < this.order ; ++i) {
  for ( j = 0 ; j < this.order ; ++j) {
   if ( matrix[i][j] != -1 ) {
    matrix[i][j] = s.elementAt( matrix[i][j] - 1 );   
 }}}
 return new Semigroup(matrix);
}
\end{lstlisting}
\end{small}

In other cases, what we need is finding all the isomorphic forms of a given
semigroup, i.e., apply all the possible permutations to a given semigroup. In
that case, we use the method \textit{permute}, which returns an array of \textit{Semigroup}
objects containing all the permutations of a given one:

\begin{small}
\begin{lstlisting}[language=Java]
public Semigroup [] permute() {
 SetS identity = new SetS( this.order );
 SetS [] permutations = identity.allPermutations() ;
 int k;
 Semigroup [] result = new Semigroup [permutations.length];
 for ( k = 0 ; k < permutations.length ; ++k) {
  result[k ] = this.permuteWith(permutations[k]);   
 }
 return result;
}
\end{lstlisting}
\end{small}

Then, a simple method to check anti-isomorphisms is given by

\begin{small}
\begin{lstlisting}[language=Java]
public Semigroup [] antiPermute () {
 return (this.transpose()).permute( );
}
\end{lstlisting}
\end{small}

In Section \ref{other_app} we will show different examples using these methods.

\subsection{Resonant decompositions}

\label{Resonant}

In Section \ref{sexpp} we explained that a resonant subalgebra can be extracted from the expanded algebra $\mathcal{G}_{S}$ when the original algebra has a certain subspace structure and the semigroup satisfy a so called resonant decomposition (see Eqs. (\ref{CDres1}-\ref{CDres2})).
As mentioned in Section \ref{descriptionLib}, for our pourpuses it will be enough to consider the case given by Eqs.(\ref{subspace_structure_}-\ref{resonant_cond}), i.e., we will study resonant conditions of the type,
\begin{align*}
S_{0} \cdot S_{0} \subset S_{0}\,,  \\
S_{0} \cdot S_{1} \subset S_{1}\,,\\
S_{1} \cdot S_{1} \subset S_{0}\,.%
\end{align*}

A previous step to check if two given subsets $S_{0}$ and $S_{1}$ satisfy the resonance condition, is
to check that their union reproduce the full semigroup, i.e. $S_{0} \cup S_{1} = S$. This is done by the
method \textit{fillTheSpace} of the \textit{SetS} class.

\begin{small}
\begin{lstlisting}[language=Java]
public static boolean fillTheSpace(SetS s1,SetS s2,int order){
 int i ;
 for ( i = 0 ; i < order ; ++i) {
  if ( ! s1.find( i+1) && ! s2.find(i +1) ){
 	 return false;
 }}
 return true;
}
\end{lstlisting}
\end{small}
The parameter \textit{order} tells the method the order of the semigroup
for which \textit{s1} and \textit{s2} must be a resonant decomposition.

The method \textit{isResonant} returns true if the two \textit{SetS} objects \textit{s0} and \textit{s1}
represent a resonant decomposition for the current \textit{Semigroup} object.
\begin{small}
\begin{lstlisting}[language=Java]
public boolean isResonant( SetS s0 , SetS s1) {
 int i,j, n0 = s0.nElements, n1 = s1.nElements ;
 if ( SetS.fillTheSpace(s0, s1, order) ) {
  for ( i = 0 ; i < n0 ; ++i ) {
   for ( j = 0 ; j < n0 ; ++j ) {
    if ( ! s0.find(this.data[s0.elementAt(i) -1]
      [s0.elementAt(j)-1] ) ) {
     return false ;
  }}}
  for ( i = 0 ; i < n0 ; ++i ){
   for ( j = 0 ; j < n1 ; ++j) {
    if ( ! s1.find( this.data[s0.elementAt(i)-1]
      [s1.elementAt(j)-1]))  {
     return false;
  }}}
  for ( i = 0 ; i < n1 ; ++i) {
   for ( j = 0 ; j < n1 ; ++j) {
    if (! s0.find( this.data[s1.elementAt(i)-1]
      [s1.elementAt(j)-1])) {
     return false;
 }}}}
 else {
  return false;
 }      
 return true;
}
\end{lstlisting}
\end{small}

Once we are able to check is a given decomposition of a semigroup is resonant,
we want to be able to look for resonant decompositions. The method
\textit{findResonances} looks for all the possible resonances of a semigroup, with
$S_{0} $ and $S_{1}$ having respectively \textit{n1} and \textit{n2} elements.

\begin{small}
\begin{lstlisting}[language=Java]
public SetS [][] findResonances( int n1, int n2 ) {
 SetS total = new SetS(this.order) ;
 SetS [] list1 = total.subSets(n1) ;
 SetS [] list2 = total.subSets(n2) ;
 SetS [][] result = null;
 SetS [][] auxiliar = null ;
 int foundResonances = 0 ;
 int i , j , k = 0;
 for ( i = 0 ; i < list1.length ; ++ i ) {
  for ( j = 0 ; j < list2.length ; ++j ) {
   if ( this.isResonant( list1[i], list2[j]) 
     && SetS.fillTheSpace( list1[i], list2[j], this.order)) {
    foundResonances = foundResonances + 1 ;
    auxiliar = result ;
    result = new SetS[foundResonances ] [2] ;
    for ( k = 0 ; k < foundResonances -1 ; ++k) {
     result[k][0] = auxiliar[k][0];
     result[k][1] = auxiliar[k][1];
    }
    result[ foundResonances - 1 ] [0] = list1[i];
    result[ foundResonances - 1] [1] = list2[j];
 }}}
 return result;
}
\end{lstlisting}
\end{small}

In case it finds any resonant decomposition, this method returns a 2
dimensional array whose element $result[i][0]$ is the $S_{0}$ and
$result[i][1]$ is $S_{1} $ for the \textit{ith} decomposition found. This
method uses the auxiliar method \textit{subSets} which returns all the subsets with $n$
element of a given \textit{SetS} object.

\begin{small}
\begin{lstlisting}[language=Java]
public SetS [] subSets( int n ) {
 SetS result = new SetS();
 return SetS.cleanDuplicates(auxSubset( this , result, n ));
}
\end{lstlisting}
\end{small}

This method is just a more convenient way to use the recursive method \textit{auxSubset}:

\begin{small}
\begin{lstlisting}[language=Java]
public SetS[] auxSubset( SetS original , SetS result, int n) {
 int i;
 SetS [] list = null;
 SetS [] totalList = null;
 if ( n == 0 ) {
  totalList = new SetS[1];
  totalList[0] = result;
  return totalList ;
 }
 SetS aux1;
 SetS aux2;
 for ( i = 0 ; i < original.nElements ; ++i ) {
  aux1= original.eraseElement(i);
  aux2= result.addElement( original.elementAt(i));
   list = auxSubset(aux1 , aux2 , n-1);
  totalList = SetS.add(list , totalList );
  if ( totalList != null ) {
 }}
 return totalList ;
}
\end{lstlisting}
\end{small}

The auxiliar method \textit{cleanDuplicates} just cleans possible duplicates in an array of \textit{SetS} objects

\begin{small}
\begin{lstlisting}[language=Java]
public static SetS[] cleanDuplicates( SetS [] lst) {
 int i, j ;
 int n = lst.length ;
 int elements = n;
 SetS [] newList ;
 for ( i = 0 ; i < n ; ++i) {
  lst[i] = SetS.sort(lst[i]);
 }
 for ( i = 0 ; i < n ; ++i) {
  for ( j = i +1 ; j < n ; ++j){
   if (lst[i] != null && lst[j] != null && lst[i].equalTo(lst[j])){
    lst[j] = null ;
    elements = elements - 1 ;   
 }}}
 newList = new SetS[elements] ;
 j = 0 ;
 for ( i = 0 ;  i < n ; ++i) {
  if ( lst[i] != null ) {
   newList[j] = lst[i];
   ++j;
 }}
 return newList ;
}
\end{lstlisting}
\end{small}

To find all the possible resonant decompositions of a given semigroup we define the method \textit{findAllResonances}

\begin{small}
\begin{lstlisting}[language=Java]
public SetS [][] findAllResonances() {
 int i , j , k;
 SetS [][] result = null;
 SetS [][] auxiliar ;
 SetS [][] intermediateResult;
 int N = 0;
 for ( i = 1 ; i < this.order ; ++i) {
  for ( j = 1 ; j < this.order ; ++j) {
   intermediateResult = this.findResonances( i, j) ;
   if ( intermediateResult != null ) {
    auxiliar = result ;
    N = N + intermediateResult.length ;
    result = new SetS[ N][2];
    for ( k = 0 ; k < N - intermediateResult.length ; ++k ) {
     result[k][0] = auxiliar[k][0];
     result[k][1] = auxiliar[k][1];
    }
    for ( k = 0 ; k < intermediateResult.length ; ++k) {
   	 result[ N - intermediateResult.length + k][0 ] = 
   	  intermediateResult[k][0];
   	 result[ N - intermediateResult.length + k][ 1 ] =
   	 intermediateResult[k][1];
 }}}}
 return result;
}
\end{lstlisting}
\end{small}

In the following Sections we will see the usefulness of these methods.

\section{Methods for S-expansions}
\label{Library_Sexpansion}

In this Section we will explain the construction of the classes which will allow us to:

\begin{itemize}
	\item Represent a semigroup of order $n$ and compute the metric $\mathbf{g}^{S}$ defined in Eq. (\ref{gs})),
	\item Represent Lie algebras in terms of its structure constants (adjoint representation) and compute the Killing-Cartan (KC) $\mathbf{g}$ metric defined in Eq. (\ref{KC_def}),
	\item Show the selector in a fancy way, by using $n$ boxes of dimension $n\times n$, so the component $K_{ab}^c $ is found in box $a$, row $b$, column $c$ (and do something similar for the structure constants of the original and expanded algebras),
	\item Obtain the S-expanded algebra, resonant subalgebra, $0_{S}$-reduced algebra and the $0_{S}$-reduction of a resonant subalgebra,
	\item Compute the KC metric of the original Lie algebra, the S-expanded algebra, the resonant subalgebra, the $0_{S}$-reduced algebra and the $0_{S}$-reduction of a resonant subalgebra (see Eqs. (\ref{KCexp}) and Section 3.2 of Ref. \cite{Andrianopoli:2013ooa})
\end{itemize}

\paragraph*{A reminder:}
\textit{For space reasons, we will only explain the main parts of those classes here. The full source code of the library is available in \cite{webJava} and its documentation can be found in \cite{wiki}.}

\subsection{Representing semigroups and Lie algebras}
\label{Rep_S_G}

In section \ref{semigroups} it was shown that an informal but useful way to represent the multiplication law of a semigroup is by means of the selectors $ K_{\alpha \beta}^\kappa $, whose definition was given in Eq. (\ref{sel1}). Thus, we define the \textit{Selector} class to represent these quantities.

\begin{small}
\begin{lstlisting}[language=Java]
public class Selector {
 int order ;
 int [][][] data ;
\end{lstlisting}
\end{small}
It has 2 variables: an integer to save the order of the semigroup and an array to save all the selectors in the semigroup as follows
\[
\mathrm{data}[a][b][c] \equiv K_{ab}^{c} \,.
\]

The methods \textit{Selector}, \textit{SetS} and \textit{get} allow to create a selector object for a semigroup of order $n$, to set and return the value of the $K_{ab}^c$ component of the selector. The method \textit{show} gives the selector by means of $n$ boxes of dimension $n\times n$, so the component $K_{ab}^c$ is found in box $a$, row $b$, column $c$. According to Eq. (\ref{sel2}), this means that box $a$ is indeed the adjoint representation of the element $\lambda_{a}$.

For example, consider the following semigroup,
\begin{equation}%
\begin{tabular}
[c]{l|llll}
& $\lambda_{1}$ & $\lambda_{2}$ & $\lambda_{3}$ & $\lambda_{4}$\\\hline
$\lambda_{1}$ & $\lambda_{1}$ & $\lambda_{2}$ & $\lambda_{3}$ & $\lambda_{4}%
$\\
$\lambda_{2}$ & $\lambda_{2}$ & $\lambda_{3}$ & $\lambda_{4}$ & $\lambda_{4}%
$\\
$\lambda_{3}$ & $\lambda_{3}$ & $\lambda_{4}$ & $\lambda_{4}$ & $\lambda_{4}%
$\\
$\lambda_{4}$ & $\lambda_{4}$ & $\lambda_{4}$ & $\lambda_{4}$ & $\lambda_{4}$%
\end{tabular}
\,.\label{Section4_sem_ex}%
\end{equation}
The following piece of code (part of the program 26 in Appendix \ref{list_ex}) ilustrates how the \textit{Selector} class works together with the method \textit{getSelector} of the \textit{Semigroup} class, to build and print a selector object for this semigroup.
\begin{small}
\begin{lstlisting}[language=Java]
 int [][] mS = {{1,2,3,4},{2,3,4,4},{3,4,4,4}, {4,4,4,4}};
 Semigroup S = new Semigroup(mS) ;
 Selector Adj = S.getSelector() ;
 Adj.show();
\end{lstlisting}
\end{small}
The output reads,
\begin{small}
\begin{verbatim}
For the considered semigroup of order m, here we print the m matrices K_{a,b}^{c}=M_{b,c}
(with a=1,...,m) which gives the adjoint representation for the elements of the semigroup.
*********
Adj [lambda_{1}] = ( K_{1,b}^{c} ) =
 1 0 0 0
 0 1 0 0
 0 0 1 0
 0 0 0 1
*********
Adj [lambda_{2}] = ( K_{2,b}^{c} ) =
 0 1 0 0
 0 0 1 0
 0 0 0 1
 0 0 0 1
*********
Adj [lambda_{3}] = ( K_{3,b}^{c} ) =
 0 0 1 0
 0 0 0 1
 0 0 0 1
 0 0 0 1
*********
Adj [lambda_{4}] = ( K_{4,b}^{c} ) =
 0 0 0 1
 0 0 0 1
 0 0 0 1
 0 0 0 1
\end{verbatim}
\end{small}

This class also contains the methods \textit{selectorMetric} and \textit{expandedMetric} which computes the metrics $\mathbf{g}^{S}$ and $\mathbf{g}^{E}$ defined in Eqs. (\ref{gs}-\ref{mario_0}). Explicit examples with them will be given in Section \ref{gen_checkings3}.

In a similar way we construct the classes \textit{SelectorReduced}, \textit{SelectorResonant} and \textit{SelectorResonantReduced}, which extend the \textit{Selector} class for the cases where the semigroup have a zero element, a resonance and both simultaneously. 

On the other hand, to represent a Lie algebra by means of its structure constants we define the \textit{StructureConstantSet} class.
\begin{small}
\begin{lstlisting}[language=Java]
public class StructureConstantSet {
 double [][][] constants ;
 int N ; 
\end{lstlisting}
\end{small}
The integer $N$ is the number of generators of the Lie algebra and, with the convention
\[
\left[  X_{i} , X_{j} \right]  = C_{ij}^{k} X_{k} \,,
\]
we choose
\[
\mathrm{constants}[i][j][k] \equiv C_{ij}^{k}\,.
\]

This class contains methods that allow to introduce the values of the structure constants and show them in a similar way as was done for the selectors.
In particular, the following method sets the values of $C_{ij}^{k} $ and $C_{ji}^{k} $.
\begin{small}
\begin{lstlisting}[language=Java]
public void setStructureConstant(int i, int j, int k, double fijk){
 constants[i][j][k] = fijk ;
 constants[j][i][k] = - fijk;
}
\end{lstlisting}
\end{small}

As the first possition of an array [ ] is usually [0], the method  \textit{setStructureConstant} must be used to introduce the non-vanishing structure constants $C_{ij}^{k}$
as:
\begin{small}
\begin{lstlisting}[language=Java]
NameOfTheAlgebra.setStructureConstant(i, j, k, value of C_{ij}^{k})
\end{lstlisting}
\end{small}
where $i,j,k=0,1,\ldots,N-1$. However, the outputs will be usually given in such a way that $i,j,k=1,\ldots,N$ . 

As an example, the following piece of code (also part of the program 26 in Appendix \ref{list_ex}) ilustrates how to introduce the structure constants of the $\mathfrak{sl}(2)$ algebra,
\begin{align}
&  \left[  X_{1},X_{2}\right]  =-2X_{3}\nonumber\\
&  \left[  X_{1},X_{3}\right]  =2X_{2}\nonumber\\
&  \left[  X_{2},X_{3}\right]  =2X_{1}\,,\label{sl2_com}%
\end{align}
in order to obtain its adjoint representation.
\begin{small}
\begin{lstlisting}[language=Java]
 sl2.setStructureConstant(0, 1, 2, -2);
 sl2.setStructureConstant(0, 2, 1, 2);
 sl2.setStructureConstant(1, 2, 0, 2);
 sl2.show();
\end{lstlisting}
\end{small}
The output reads,
\begin{small}
\begin{verbatim}
For the considered Lie algebra of dimension n, we print the n matrices C_{ij}^{k}=M_{jk}
(with i=1,...,n) which gives the adjoint representation for the elements of the algebra.
*********
Adj [ X_{1} ] = ( C_{1,j}^{k} ) =
 0.0  0.0  0.0
 0.0  0.0 -2.0
 0.0  2.0  0.0
*****
*********
Adj [ X_{2} ] = ( C_{2,j}^{k} ) =
 0.0  0.0  2.0
 0.0  0.0  0.0
 2.0  0.0  0.0
*****
*********
Adj [ X_{3} ] = ( C_{3,j}^{k} ) =
 0.0 -2.0  0.0
-2.0  0.0  0.0
 0.0  0.0  0.0
\end{verbatim}
\end{small}

Several simple operations can be perfomed by the methods of the \textit{StructureConstantSet} class. For example, the following one computes the Killing-Cartán metric of a given Lie algebra.
\begin{small}
\begin{lstlisting}[language=Java]
public Matrix cartanKillingMetric(){
 int i, j , k, l;
 double sum = 0 ;
 double [][] metric = new double[N][N] ;
 for ( i = 0 ; i < N ; ++i) {
  for ( j = 0 ; j < N ; ++j ) {
   sum = 0 ;
   for ( k = 0 ; k < N ; ++k ) {
    for ( l = 0 ; l < N ; ++l ) {
     sum = sum + this.structureConstant(i,k,l) 
      * this.structureConstant(j,l,k);
   }}
   metric[i][j] = sum ;
 }}
 return new Matrix( metric );
}
\end{lstlisting}
\end{small}
Explicit examples using this and other methods included in the classes \textit{Selector}, \textit{StructureConstantSet} will be given in Section \ref{gen_checkings3}.

\subsection{The S-expanded algebra}
\label{Rep_GS}

In double index notation, the Lie bracket of the S-expanded algebra is written here as
\[
\left[  X_{\left(i,a\right)  },X_{\left(j,b\right)}\right]
=C_{ij}^{k}K_{ab}^{c}X_{\left(k,c\right)}\,.
\]
To use this double index notation internally in our library, we create the class \textit{StructureConstantSetExpanded}.

\begin{small}
\begin{lstlisting}[language=Java]
public class StructureConstantSetExpanded {
 int n,m ;
 double [][][][][][] data ;
\end{lstlisting}
\end{small}

Thus, an object of this class has information about the dimension $n$ of the original
Lie algebra, the order $m$ of the semigroup used to perform the S-expansion and uses the 6-dimensional array \textit{data} to save the structure constants, in such a way that
\[
\mathrm{data}[i][a][j][b][k][c] \equiv C_{(i,a)(j,b)}^{(k,c)}\,.
\]

To get the S-expanded algebra we must follow the next steps:

\begin{enumerate}
\item Create a \textit{StructureConstantSet} object to store the original Lie algebra,

\item Create a \textit{Semigroup} object \textit{S} to store the semigroup which we want to use
for the S-expansion,

\item Use the method \textit{getExpandedStructureConstant} to perform the S-expansion of the Lie algebra with the semigroup object,

\item Use the method \textit{showCommut} to get the non vanishing commutators of the expanded algebra $\mathcal{G}_{S}$,

\item Use the method \textit{showSC} to get the non vanishing structure constants of $\mathcal{G}_{S}$,

\item Use the method \textit{cartanKillingMetric} to compute the KC metric of $\mathcal{G}_{S}$.

\end{enumerate}

A piece of code performing the steps above for the algebra and the semigroup given by Eqs. (\ref{sl2_com}) and (\ref{Section4_sem_ex}), would be
\begin{small}
\begin{lstlisting}[language=Java]
 StructureConstantSet sl2 = new StructureConstantSet(3) ;
 sl2.setStructureConstant(0, 1, 2,-2) ;
 sl2.setStructureConstant(0, 2, 1, 2) ;
 sl2.setStructureConstant(1, 2, 0, 2) ;
 int [][] SemigroupTable = {{1,2,3,4},{2,3,4,4},{3,4,4,4}, {4,4,4,4}};
 Semigroup S = new Semigroup(SemigroupTable) ;
 StructureConstantSetExpanded Gexp = S.getExpandedStructureConstant(sl2);
 Gexp.showCommut();
 Gexp.showSC();
 Matrix metricaGexp = Gexp.cartanKillingMetric();
 metricaGexp.print(0,0);
\end{lstlisting}
\end{small}
Further examples with sample outputs will be given in Section \ref{gen_checkings3}.

\subsection{The resonant subalgebra}
\label{Rep_GSRes}

To get a resonant subalgebra of a S-expanded algebra, we define the
\textit{StructureConstantSetExpandedResonant} class, which is a child of the \textit{StructureConstantSetExpanded} class. 

\begin{small}
\begin{lstlisting}[language=Java]
public class StructureConstantSetExpandedResonant extends
  StructureConstantSetExpanded {
 SetS S0, S1, V0, V1;
\end{lstlisting}
\end{small}

Here \textit{S0} and \textit{S1} represent the resonant decomposition of the semigroup, whilst
\textit{V0} and \textit{V1} give the graded decomposition of the Lie algebra. To get the
resonant subalgebra of an S-expanded algebra we must

\begin{enumerate}
\item Get the correspondent S-expanded algebra, following the steps in the previous section,

\item Introduce the resonant decomposition $S_{0}$ and $S_{1}$ with the method \textit{SetS},

\item Introduce the graded decomposition $V_{0} $ and $V_{1} $ also with the method \textit{SetS},

\item Use all these objects to create a \textit{StructureConstantSetExpandedResonant} object,

\item Use methods \textit{showCommutRes}, \textit{showSCRes} and \textit{cartanKillingMetricPretty} to obtain the non vanishing commutators, structure constants and KC metric of the resonant subalgebra.

\end{enumerate}

For example, the algebra considered in Eq. (\ref{sl2_com}) has a subspace decomposition given by $\mathfrak{sl}(2)=V_{0}\oplus V_{1}$\,, where
\[
V_{0} = \{ X_{1} \} \text{  and  } V_{1} = \{ X_{2} , X_{3} \} \,.
\]
Considering again the semigroup given in Eq. (\ref{Section4_sem_ex}), we see that admits a decomposition $S=S_{0}\cup S_{1}$ with
\[
S_{0} = \{1,3,4\} \text{  and  } S_{1} = \{ 2,4 \}  \,,
\]
which is resonant, i.e., it satisfies Eq. (\ref{resonant_cond}).
Thus, the following piece of code perform the S-expansion of $\mathfrak{sl}(2)$ and it calculates the resonant subalgebra.
\begin{small}
\begin{lstlisting}[language=Java]
 StructureConstantSet sl2 = new StructureConstantSet(3) ;
 sl2.setStructureConstant(0, 1, 2,-2) ;
 sl2.setStructureConstant(0, 2, 1, 2) ;
 sl2.setStructureConstant(1, 2, 0, 2) ;
 int [][] SemigroupTable = {{1,2,3,4},{2,3,4,4},{3,4,4,4}, {4,4,4,4}};
 Semigroup S = new Semigroup(SemigroupTable) ;
 int[] mS0 = {1,3,4} ;
 SetS S0 = new SetS(mS0) ;
 int[] mS1 = {2,4} ;
 SetS S1 = new SetS(mS1) ;
 int[] mV0 = {1};
 SetS V0 = new SetS(mV0);
 int[] mV1 = {2,3};
 SetS V1 = new SetS(mV1);
 StructureConstantSetExpanded Gexp = S.getExpandedStructureConstant(sl2);
 StructureConstantSetExpandedResonant ResGexp = 
  new StructureConstantSetExpandedResonant(Gexp.data,S0,S1,V0,V1) ;
 ResGexp.showCommutRes();
 ResGexp.showSCRes();
 Matrix metricaResGexp = ResGexp.cartanKillingMetricPretty();
 metricaResGexp.print(0,0);
\end{lstlisting}
\end{small}

Other examples with samples of their outputs will be described in detail in Section \ref{gen_checkings3}.

\subsection{S-expanded algebra followed by a $0_{S}$-reduction}
\label{Rep_GSRed}

To perform the reduction by the zero element of a S-expanded Lie algebra we
define the \textit{StructureConstantSetExpandedReduced} class, which is a child of the
\textit{StructureConstantSetExpanded} class. 

\begin{small}
\begin{lstlisting}[language=Java]
public class StructureConstantSetExpandedReduced extends
 StructureConstantSetExpanded {
  int zero ;
\end{lstlisting}
\end{small}

Respect to the \textit{StructureConstantSetExpanded} class, it only adds an integer variable to save the zero element.

What we need to perform the reduction by zero of a S-expanded algebra we have to

\begin{enumerate}
\item Get the S-expanded algebra in a \textit{StructureConstantSetExpanded},

\item Use it to create a \textit{StructureConstantSetExpandedReduced},

\item Use other methods in the class to obtain the non vanishing commutators, structure constants and KC metric of the reduced algebra.

\end{enumerate}

For example, the semigroup given by Eq. (\ref{Section4_sem_ex}) has $\lambda_{4} $ as the zero element. Thus, the following piece of code calculates the $0_{S}$-reduction of the S-expansion of the algebra $\mathfrak{sl}(2)$.
\begin{small}
\begin{lstlisting}[language=Java]
 StructureConstantSet sl2 = new StructureConstantSet(3) ;
 sl2.setStructureConstant(0, 1, 2,-2) ;
 sl2.setStructureConstant(0, 2, 1, 2) ;
 sl2.setStructureConstant(1, 2, 0, 2) ;
 int [][] SemigroupTable = {{1,2,3,4},{2,3,4,4},{3,4,4,4}, {4,4,4,4}};
 Semigroup S = new Semigroup(SemigroupTable);
 StructureConstantSetExpandedReduced RedGexp = 
  new StructureConstantSetExpandedReduced
  (S.getExpandedStructureConstant(sl2).data,4); 
 RedGexp.showCommutRed();
 RedGexp.showSCRed();
 Matrix metricaRedGexp = RedGexp.cartanKillingMetricPretty();
 metricaRedGexp.print(0,0);
\end{lstlisting}
\end{small}

\subsection{Resonant subalgebra followed by a $0_{S}$-reduction} 
\label{Rep_GSResRed}

We define the class \textit{StructureConstantSetExpandedResonantReduced}, which is a child of the class \textit{StructureConstantSetExpandedResonant}.

\begin{small}
\begin{lstlisting}[language=Java]
public class StructureConstantSetExpandedResonantReduced
 extends StructureConstantSetExpandedResonant {
 int zero;
\end{lstlisting}
\end{small}

We just add an integer to save the zero element. Its use is analogous to that of the \textit{StructureConstantSetExpandedResonant} described in Section \ref{Rep_GSRes}. Explicit examples using this class and its methods are given in Section \ref{gen_checkings3}.

\section{Applications}

\label{other_app}

Here we will describe most of the programs listed in Appendix \ref{list_ex} in order to show explicitly the kind of calculations that can be done with the methods described in the sections \ref{generating} and \ref{Library_Sexpansion}. Mainly, we will explain the ones that allow us to:
\begin{itemize}
	\item Check associativity and commutativity, find the zero element and resonances for any given multiplication table,
	\item Apply permutations and find isomorphisms between any set of semigroups,
	\item Perform S-expansions with any given semigroup and, if the semigroup fulfills the necessary conditions, to find the resonant subalgebra $\mathcal{G}_{S,R}$, the reduced algebra $\mathcal{G}_{S,\text{red}}$ and the reduction of the resonant subalgebra $\mathcal{G}_{S,R,\text{red}}$,
	\item Identify S-expansions preserving semisimplicity.
\end{itemize}

With these examples the user can easily create new programs to perform his own calculations. 
Along this section, we will refer to the example programs by their Appendix-\ref{list_ex} number. When the output is short, the results are direclty printed to screen.
Longer outputs are printed in \textit{.txt}-files\footnote{For a good visualization, those files must opened with Notepad++.} in the folder \textit{``./Output\_examples/''}, with the naming convention: \textit{Output\_ + name of the program}. 
In addition, most of the outputs can be found in the file \textit{Output\_examples.zip} available in \cite{webJava}.
Finally, we remind that a brief description about the general structure of these programs can be found in Section \ref{descriptionLib}.

\subsection{Examples of associativity and commutativity}
\label{asso_com}

The first two programs listed in Appendix \ref{list_ex} allow to check the associativity and commutativity of the following multiplication tables:

\[%
\begin{tabular}
[c]{l|lll}%
$S_{\text{ex1}}$ & $\lambda_{1}$ & $\lambda_{2}$ & $\lambda_{3}$\\\hline
$\lambda_{1}$ & $\lambda_{1}$ & $\lambda_{2}$ & $\lambda_{3}$\\
$\lambda_{2}$ & $\lambda_{2}$ & $\lambda_{1}$ & $\lambda_{2}$\\
$\lambda_{3}$ & $\lambda_{3}$ & $\lambda_{2}$ & $\lambda_{1}$%
\end{tabular}
\,,\ \
\begin{tabular}
[c]{l|lll}%
$S_{\text{ex2}}$ & $\lambda_{1}$ & $\lambda_{2}$ & $\lambda_{3}$\\\hline
$\lambda_{1}$ & $\lambda_{1}$ & $\lambda_{1}$ & $\lambda_{1}$\\
$\lambda_{2}$ & $\lambda_{1}$ & $\lambda_{2}$ & $\lambda_{3}$\\
$\lambda_{3}$ & $\lambda_{1}$ & $\lambda_{1}$ & $\lambda_{1}$%
\end{tabular}
\,,\ \
\begin{tabular}
[c]{l|lll}%
$S_{\text{ex3}}$ & $\lambda_{1}$ & $\lambda_{2}$ & $\lambda_{3}$\\\hline
$\lambda_{1}$ & $\lambda_{1}$ & $\lambda_{1}$ & $\lambda_{3}$\\
$\lambda_{2}$ & $\lambda_{1}$ & $\lambda_{2}$ & $\lambda_{3}$\\
$\lambda_{3}$ & $\lambda_{3}$ & $\lambda_{3}$ & $\lambda_{1}$%
\end{tabular}
\,,
\]%

\[%
\begin{tabular}
[c]{l|llll}%
$S_{\text{ex4}}$ & $\lambda_{1}$ & $\lambda_{2}$ & $\lambda_{3}$ &
$\lambda_{4}$\\\hline
$\lambda_{1}$ & $\lambda_{1}$ & $\lambda_{2}$ & $\lambda_{3}$ & $\lambda_{4}%
$\\
$\lambda_{2}$ & $\lambda_{2}$ & $\lambda_{3}$ & $\lambda_{4}$ & $\lambda_{1}%
$\\
$\lambda_{3}$ & $\lambda_{3}$ & $\lambda_{4}$ & $\lambda_{1}$ & $\lambda_{2}%
$\\
$\lambda_{4}$ & $\lambda_{4}$ & $\lambda_{1}$ & $\lambda_{2}$ & $\lambda_{3}$%
\end{tabular}
\,,\ \
\begin{tabular}
[c]{l|lllll}%
$S_{\text{ex5}}$ & $\lambda_{1}$ & $\lambda_{2}$ & $\lambda_{3}$ &
$\lambda_{4}$ & $\lambda_{5}$\\\hline
$\lambda_{1}$ & $\lambda_{5}$ & $\lambda_{4}$ & $\lambda_{2}$ & $\lambda_{1}$
& $\lambda_{3}$\\
$\lambda_{2}$ & $\lambda_{4}$ & $\lambda_{3}$ & $\lambda_{5}$ & $\lambda_{2}$
& $\lambda_{1}$\\
$\lambda_{3}$ & $\lambda_{2}$ & $\lambda_{5}$ & $\lambda_{1}$ & $\lambda_{3}$
& $\lambda_{4}$\\
$\lambda_{4}$ & $\lambda_{1}$ & $\lambda_{2}$ & $\lambda_{3}$ & $\lambda_{4}$
& $\lambda_{5}$\\
$\lambda_{5}$ & $\lambda_{3}$ & $\lambda_{1}$ & $\lambda_{4}$ & $\lambda_{5}$
& $\lambda_{2}$%
\end{tabular}
\,,
\]

\[%
\begin{tabular}
[c]{l|llllllllll}%
$S_{\text{ex6}}$ & $\lambda_{1}$ & $\lambda_{2}$ & $\lambda_{3}$ &
$\lambda_{4}$ & $\lambda_{5}$ & $\lambda_{6}$ & $\lambda_{7}$ & $\lambda_{8}$
& $\lambda_{9}$ & $\lambda_{10}$\\\hline
$\lambda_{1}$ & $\lambda_{1}$ & $\lambda_{1}$ & $\lambda_{1}$ & $\lambda_{1}$
& $\lambda_{5}$ & $\lambda_{5}$ & $\lambda_{1}$ & $\lambda_{1}$ & $\lambda
_{5}$ & $\lambda_{5}$\\
$\lambda_{2}$ & $\lambda_{1}$ & $\lambda_{2}$ & $\lambda_{1}$ & $\lambda_{1}$
& $\lambda_{5}$ & $\lambda_{5}$ & $\lambda_{1}$ & $\lambda_{1}$ & $\lambda
_{5}$ & $\lambda_{5}$\\
$\lambda_{3}$ & $\lambda_{3}$ & $\lambda_{3}$ & $\lambda_{3}$ & $\lambda_{3}$
& $\lambda_{6}$ & $\lambda_{6}$ & $\lambda_{3}$ & $\lambda_{3}$ & $\lambda
_{6}$ & $\lambda_{6}$\\
$\lambda_{4}$ & $\lambda_{3}$ & $\lambda_{3}$ & $\lambda_{3}$ & $\lambda_{4}$
& $\lambda_{6}$ & $\lambda_{6}$ & $\lambda_{3}$ & $\lambda_{3}$ & $\lambda
_{6}$ & $\lambda_{6}$\\
$\lambda_{5}$ & $\lambda_{1}$ & $\lambda_{1}$ & $\lambda_{1}$ & $\lambda_{1}$
& $\lambda_{5}$ & $\lambda_{5}$ & $\lambda_{5}$ & $\lambda_{5}$ & $\lambda
_{1}$ & $\lambda_{1}$\\
$\lambda_{6}$ & $\lambda_{3}$ & $\lambda_{3}$ & $\lambda_{3}$ & $\lambda_{3}$
& $\lambda_{6}$ & $\lambda_{6}$ & $\lambda_{6}$ & $\lambda_{6}$ & $\lambda
_{3}$ & $\lambda_{3}$\\
$\lambda_{7}$ & $\lambda_{1}$ & $\lambda_{1}$ & $\lambda_{3}$ & $\lambda_{3}$
& $\lambda_{5}$ & $\lambda_{6}$ & $\lambda_{7}$ & $\lambda_{8}$ & $\lambda
_{9}$ & $\lambda_{10}$\\
$\lambda_{8}$ & $\lambda_{1}$ & $\lambda_{1}$ & $\lambda_{3}$ & $\lambda_{3}$
& $\lambda_{5}$ & $\lambda_{6}$ & $\lambda_{8}$ & $\lambda_{7}$ &
$\lambda_{10}$ & $\lambda_{9}$\\
$\lambda_{9}$ & $\lambda_{3}$ & $\lambda_{3}$ & $\lambda_{1}$ & $\lambda_{1}$
& $\lambda_{6}$ & $\lambda_{5}$ & $\lambda_{9}$ & $\lambda_{10}$ &
$\lambda_{8}$ & $\lambda_{7}$\\
$\lambda_{10}$ & $\lambda_{3}$ & $\lambda_{3}$ & $\lambda_{1}$ & $\lambda_{1}$
& $\lambda_{6}$ & $\lambda_{5}$ & $\lambda_{10}$ & $\lambda_{9}$ &
$\lambda_{7}$ & $\lambda_{8}$%
\end{tabular}\,.
\]%
\bigskip

The main parts of the program 1 are described in what follows.

\begin{small}
\begin{lstlisting}[language=Java]
package examples;
import sexpansion.Semigroup;
public class I_associative_checking {
 public static void main(String[] args) {
 // Manually enter the semigroups as matrices
 int[][] ms_ex1 = {{1,2,3},{2,1,2},{3,2,1}};
 // and do the same for the other tables.
 // Next, we use these tables to define Semigroup objects:
 Semigroup s_ex1 = new Semigroup( ms_ex1) ;
 // and define s_ex2,..., s_ex6 in the same way.
 // Now, the method isAssociative can be used as follows:
 if (s_ex1.isAssociative()) {
  System.out.println("The semigroup S_ex1 is associative");}
 else {
  System.out.println("The semigroup S_ex1 is not associative");}
 // and proceed similarly with the other tables.
}}
\end{lstlisting}
\end{small}
The output reads
\begin{small}
\begin{verbatim}
The semigroup S_ex1 is not associative,
The semigroup S_ex2 is associative,
The semigroup S_ex3 is associative,
The semigroup S_ex4 is associative,
The semigroup S_ex5 is associative,
The semigroup S_ex6 is associative.
\end{verbatim}
\end{small}

Program 2 is defined in a similar way.
\begin{small}
\begin{lstlisting}[language=Java]
package examples;
import sexpansion.Semigroup;
public class I_commutative_checking {
 public static void main(String[] args) {
\end{lstlisting}
\end{small}
After introducing the tables, in the same way as was done for the program 1, 
we use the method \textit{isCommutative} as follows,

\begin{small}
\begin{lstlisting}[language=Java]
 if ( s_ex1.isCommutative()) {
  System.out.println("S_ex1 is commutative");}
 else {
  System.out.println("S_ex1 is not commutative");} 
 // and repeat the same for the other tables.
}}
\end{lstlisting}
\end{small}
Thus, we have the following result
\begin{small}
\begin{verbatim}
S_ex1 is commutative,
S_ex2 is not commutative,
S_ex3 is commutative,
S_ex4 is commutative,
S_ex5 is commutative,
S_ex6 is not commutative.
\end{verbatim}
\end{small}
In particular, the semigroup $S_{\text{ex6}}$ is an example of a non commutative semigroup of order 10 that has been used in Ref. \cite{S10}.

Another example is given by the programs 3, which selects the abelian semigroups from the lists of all non isomorphic semigroups \textit{sem.n} (with $n=2,\ldots,6$). The outputs are printed in the folder \textit{``./Output\_examples/''}, in text files named with the convention explained at the begining of this section.

\subsection{Examples with zero element and resonances}
\label{gen_checkings}

Consider the following multiplication tables:
\begin{align}
&
\begin{tabular}
[c]{l|llll}%
$S_{E}^{\left(  2\right)  }$ & $\lambda_{0}$ & $\lambda_{1}$ & $\lambda_{2}$ &
$\lambda_{3}$\\\hline
$\lambda_{0}$ & $\lambda_{0}$ & $\lambda_{1}$ & $\lambda_{2}$ & $\lambda_{3}%
$\\
$\lambda_{1}$ & $\lambda_{1}$ & $\lambda_{2}$ & $\lambda_{3}$ & $\lambda_{3}%
$\\
$\lambda_{2}$ & $\lambda_{2}$ & $\lambda_{3}$ & $\lambda_{3}$ & $\lambda_{3}%
$\\
$\lambda_{3}$ & $\lambda_{3}$ & $\lambda_{3}$ & $\lambda_{3}$ & $\lambda_{3}$%
\end{tabular}
\,,\ \
\begin{tabular}
[c]{l|llll}%
$S_{K}^{\left(  3\right)  }$ & $\lambda_{1}$ & $\lambda_{2}$ & $\lambda_{3}$ &
$\lambda_{4}$\\\hline
$\lambda_{1}$ & $\lambda_{4}$ & $\lambda_{4}$ & $\lambda_{1}$ & $\lambda_{4}%
$\\
$\lambda_{2}$ & $\lambda_{4}$ & $\lambda_{2}$ & $\lambda_{2}$ & $\lambda_{4}%
$\\
$\lambda_{3}$ & $\lambda_{1}$ & $\lambda_{2}$ & $\lambda_{3}$ & $\lambda_{4}%
$\\
$\lambda_{4}$ & $\lambda_{4}$ & $\lambda_{4}$ & $\lambda_{4}$ & $\lambda_{4}$%
\end{tabular}
\,,\ \
\begin{tabular}
[c]{l|llll}%
$S_{N1}$ & $\lambda_{1}$ & $\lambda_{2}$ & $\lambda_{3}$ & $\lambda_{4}%
$\\\hline
$\lambda_{1}$ & $\lambda_{4}$ & $\lambda_{4}$ & $\lambda_{1}$ & $\lambda_{4}%
$\\
$\lambda_{2}$ & $\lambda_{4}$ & $\lambda_{2}$ & $\lambda_{4}$ & $\lambda_{4}%
$\\
$\lambda_{3}$ & $\lambda_{1}$ & $\lambda_{4}$ & $\lambda_{3}$ & $\lambda_{4}%
$\\
$\lambda_{4}$ & $\lambda_{4}$ & $\lambda_{4}$ & $\lambda_{4}$ & $\lambda_{4}$%
\end{tabular}
\,,\nonumber\\
&
\begin{tabular}
[c]{l|llll}%
$S_{N2}$ & $\lambda_{1}$ & $\lambda_{2}$ & $\lambda_{3}$ & $\lambda_{4}%
$\\\hline
$\lambda_{1}$ & $\lambda_{2}$ & $\lambda_{3}$ & $\lambda_{4}$ & $\lambda_{4}%
$\\
$\lambda_{2}$ & $\lambda_{3}$ & $\lambda_{4}$ & $\lambda_{4}$ & $\lambda_{4}%
$\\
$\lambda_{3}$ & $\lambda_{4}$ & $\lambda_{4}$ & $\lambda_{4}$ & $\lambda_{4}%
$\\
$\lambda_{4}$ & $\lambda_{4}$ & $\lambda_{4}$ & $\lambda_{4}$ & $\lambda_{4}$%
\end{tabular}
\,,\ \
\begin{tabular}
[c]{l|llll}%
$S_{N3}$ & $\lambda_{1}$ & $\lambda_{2}$ & $\lambda_{3}$ & $\lambda_{4}%
$\\\hline
$\lambda_{1}$ & $\lambda_{4}$ & $\lambda_{1}$ & $\lambda_{4}$ & $\lambda_{4}%
$\\
$\lambda_{2}$ & $\lambda_{1}$ & $\lambda_{2}$ & $\lambda_{3}$ & $\lambda_{4}%
$\\
$\lambda_{3}$ & $\lambda_{4}$ & $\lambda_{3}$ & $\lambda_{4}$ & $\lambda_{4}%
$\\
$\lambda_{4}$ & $\lambda_{4}$ & $\lambda_{4}$ & $\lambda_{4}$ & $\lambda_{4}$%
\end{tabular}
\,,\ \
\begin{tabular}
[c]{l|llll}%
$S_{S3}$ & $\lambda_{0}$ & $\lambda_{1}$ & $\lambda_{2}$ & $\lambda_{3}%
$\\\hline
$\lambda_{0}$ & $\lambda_{2}$ & $\lambda_{3}$ & $\lambda_{0}$ & $\lambda_{3}%
$\\
$\lambda_{1}$ & $\lambda_{3}$ & $\lambda_{1}$ & $\lambda_{3}$ & $\lambda_{3}%
$\\
$\lambda_{2}$ & $\lambda_{0}$ & $\lambda_{3}$ & $\lambda_{2}$ & $\lambda_{3}%
$\\
$\lambda_{3}$ & $\lambda_{3}$ & $\lambda_{3}$ & $\lambda_{3}$ & $\lambda_{3}$%
\end{tabular}
\,,\nonumber\\
&
\begin{tabular}
[c]{l|lll}%
$S_{S2}$ & $\lambda_{0}$ & $\lambda_{1}$ & $\lambda_{2}$\\\hline
$\lambda_{0}$ & $\lambda_{0}$ & $\lambda_{1}$ & $\lambda_{2}$\\
$\lambda_{1}$ & $\lambda_{1}$ & $\lambda_{2}$ & $\lambda_{1}$\\
$\lambda_{2}$ & $\lambda_{2}$ & $\lambda_{1}$ & $\lambda_{2}$%
\end{tabular}
\,,\ \
\begin{tabular}
[c]{l|llll}%
$S_{M}^{\left(  3\right)  }$ & $\lambda_{0}$ & $\lambda_{1}$ & $\lambda_{2}$ &
$\lambda_{3}$\\\hline
$\lambda_{0}$ & $\lambda_{0}$ & $\lambda_{1}$ & $\lambda_{2}$ & $\lambda_{3}%
$\\
$\lambda_{1}$ & $\lambda_{1}$ & $\lambda_{2}$ & $\lambda_{3}$ & $\lambda_{0}%
$\\
$\lambda_{2}$ & $\lambda_{2}$ & $\lambda_{3}$ & $\lambda_{0}$ & $\lambda_{1}%
$\\
$\lambda_{3}$ & $\lambda_{3}$ & $\lambda_{0}$ & $\lambda_{1}$ & $\lambda_{2}$%
\end{tabular}
\,,\ \
\begin{tabular}
[c]{l|lllll}%
$S_{M}^{\left(  4\right)  }$ & $\lambda_{0}$ & $\lambda_{1}$ & $\lambda_{2}$ &
$\lambda_{3}$ & $\lambda_{4}$\\\hline
$\lambda_{0}$ & $\lambda_{0}$ & $\lambda_{1}$ & $\lambda_{2}$ & $\lambda_{3}$
& $\lambda_{4}$\\
$\lambda_{1}$ & $\lambda_{1}$ & $\lambda_{2}$ & $\lambda_{3}$ & $\lambda_{4}$
& $\lambda_{1}$\\
$\lambda_{2}$ & $\lambda_{2}$ & $\lambda_{3}$ & $\lambda_{4}$ & $\lambda_{1}$
& $\lambda_{2}$\\
$\lambda_{3}$ & $\lambda_{3}$ & $\lambda_{4}$ & $\lambda_{1}$ & $\lambda_{2}$
& $\lambda_{3}$\\
$\lambda_{4}$ & $\lambda_{4}$ & $\lambda_{1}$ & $\lambda_{2}$ & $\lambda_{3}$
& $\lambda_{4}$%
\end{tabular}
\,.\label{Ex_B_CS}%
\end{align}

With a simple program, like the ones described in the previous Section \ref{asso_com}, it can be checked that the tables above actually represent abelian semigroups. Indeed, they have been used in different works related with S-expansions. First, the semigroups $S_{E}^{\left(  2\right)  }$, $S_{K}^{\left(  3\right)  }$,
$S_{N1}$, $S_{N2}$ and $S_{N3}$ were used in Ref. \cite{Caroca:2011qs} to obtain some Bianchi as an
S-expansion from the 2-dimensional solvable algebra $\left[  X_{1},X_{2}\right]
=X_{1}$. 
Then, semigroup $S_{S3}$ and $S_{S2}$ were constructed in Ref. \cite{Diaz:2012zza} to show that the semisimple version of the Maxwell algebra \cite{Soroka:2006aj} can
be obtained as an S-expansion of the AdS algebra, $\mathfrak{so}\left(D-1,2\right)$. Later, in Ref. \cite{Salgado:2014qqa}, the semigroup $S_{S2}$ was renamed as $S_{M}^{\left(
2\right)  }$ and extended to a family of semigroups $S_{M}^{\left(  N\right)  }$, which lead to a generalization of the Maxwell algebras.

Now we will show explicitly how the methods of our library can be used to find the zero element and all resonant decompositions (of the type (\ref{resonant_cond})) for these semigroups. 
First, the program 19 introduces those semigroups as tables and then as semigroup objects, in the same way explained in Section \ref{asso_com} for the program 1. 
\begin{small}
\begin{lstlisting}[language=Java]
package examples;
import sexpansion.Semigroup;
public class II_findzero_ex {
 public static void main(String[] args) {
 // Manually enter the semigroups table
\end{lstlisting}
\end{small}
The only important issue is that the element $\lambda_{0}$ in the multiplication tables $S_{E}^{\left(  2\right)  }$, $S_{S3}$, $S_{S2}$, $S_{M}^{\left(  3\right)  }$, $S_{M}^{\left(  4\right)  }$ cannot be represented by the number $0$, because the library recognizes only positive interger numbers to represent semigroup elements. For example, the matrix associated to $S_{S2}$ must be introduced as 
\begin{small}
\begin{lstlisting}[language=Java]
  int[][] ms_ex1 = {{1,2,3},{2,3,2},{3,2,3}};
\end{lstlisting}
\end{small}
where we have identified $\lambda_{0}$ with $1$, $\lambda_{1}$ with $2$ and $\lambda_{2}$ with $3$. And a similar change must be applied to introduce the other semigroups having the element $\lambda_{0}$.

Then, the program 19 uses the method \textit{findZero} of the class \textit{Semigroup} to define an interger for each semigroup to save its zero element. It also defines the interger \textit{nozero} that is used when a given semigroup has no zero element.
\begin{small}
\begin{lstlisting}[language=Java]
  int zero_S_E2 = S_E2.findZero();
  // and do the same for the others.
int nozero = -1;
\end{lstlisting}
\end{small}
The following piece of code is used to obtain the results.
\begin{small}
\begin{lstlisting}[language=Java]
  if (zero_S_E2 == nozero){
  System.out.println("The semigroup S_E2 has no zero element");}
  else {
  System.out.println("The zero element of S_E2 is " +zero_S_E2);}
  // and repeat for the others. 
}}
\end{lstlisting}
\end{small}
Thus, the output reads
\begin{small}
\begin{verbatim}
The zero element of S_E2 is 4,
The zero element of S_K3 is 4,
The zero element of S_N1 is 4,
The zero element of S_N2 is 4,
The zero element of S_N3 is 4,
The zero element of S_S3 is 4,
The semigroup S_S2 has no zero element,
The semigroup S_M3 has no zero element,
The semigroup S_M4 has no zero element.
\end{verbatim}
\end{small}

\bigskip

On the other hand, the program 15 finds\footnote{The program 16 does the same as the program 15, but prints the result in the folder \textit{``./Output\_examples/''}.} the resonances of the semigroups given in (\ref{Ex_B_CS}).
After introducing the semigroup tables, it defines the following variables: two auxiliar \textit{SetS} objects \textit{S0} and \textit{S1}, the intergers $j$ and \textit{nResonances} to count the different resonances that a given semigroup may have.
\begin{small}
\begin{lstlisting}[language=Java]
package examples;
import sexpansion.Semigroup;
import sexpansion.SetS;
public class II_findresonances_ex_console {
 public static void main(String[] args) {
  // Manually enter the semigroups table
  SetS S0;
  SetS S1;
  int j, nResonances = 0;
\end{lstlisting}
\end{small}
Then it defines a 2-dimensional array for each semigroup, whose elements are \textit{SetS} objects.
\begin{small}
\begin{lstlisting}[language=Java]
  SetS[][] rS_E2;
  ... // Similarly define rS_N1, ..., rS_M4 for each semigroup.
\end{lstlisting}
\end{small}
To find the resonant decompositions of each semigroup, the method \textit{findAllResonances} is used as follows.
\begin{small}
\begin{lstlisting}[language=Java]
  rS_E2 = S_E2.findAllResonances();
  ... // And similarly define rS_K3, ..., rS_M4 for the other semigroups.
\end{lstlisting}
\end{small}
This means that for each element in the array, let us call it $rS\_name$, $rS\_name$[j][0] and $rS\_name$[j][1] will play respectively the role of $S_{0}$ 
and $S_{1}$ in the $j$th resonant decomposition for the semigroup $S\_name$. This can be explicitly seen in the following piece of code, which is used to print the result.
\begin{small}
\begin{lstlisting}[language=Java]
  if (rS_E2 != null) {
   System.out.println("The semigroup S_E2 has "+rS_E2.length+ " resonances:");
   for ( j = 0 ; j < rS_E2.length ; ++j ) {
    nResonances = nResonances + 1;
    S0 = rS_E2[j][0] ;
    S1 = rS_E2[j][1] ;
    System.out.println("Resonance #" +nResonances);
    System.out.print("S0: ");
    S0.show();
    System.out.print("S1: " ) ;
    S1.show();
   }
   nResonances = 0;
  }
  ... // And repeat a similar code for rS_K3, ..., rS_M4.
}}
\end{lstlisting}
\end{small}
A sample output looks as follows.
\begin{small}
\begin{verbatim}
...
The semigroup S_N3 has 5 resonances:
Resonance #1
S0: 2 4
S1: 1 3 4
Resonance #2
S0: 2 3 4
S1: 1 4
Resonance #3
S0: 1 2 4
S1: 3 4
Resonance #4
S0: 2 3 4
S1: 1 3 4
Resonance #5
S0: 1 2 4
S1: 1 3 4
The semigroup S_S3 has 2 resonances:
...
\end{verbatim}
\end{small}

Remarkably, a semigroup having more than one resonant decomposition might lead, through the S-expansion procedure, to different non-isomorphic expanded Lie algebras.
This is the case of the semigroup $S_{N3}$, which according to the summary given in table 5 and Eqs. 34 and 36 of Ref. \cite{Caroca:2011qs}, allows to obtain the Type III and V Bianchi algebras with the resonant conditions:
\begin{align}
S_{0}=\{\lambda_{2},\lambda_{4}\}\,, \ \ S_{1}=\{\lambda_{1},\lambda_{3},\lambda
_{4}\}\text{.} \label{ch3}%
\end{align}
and
\begin{align}
S_{0}=\{\lambda_{2},\lambda_{3},\lambda_{4}\}\,, \ \ S_{1}=\{\lambda_{1},\lambda
_{4}\}\text{.} \label{ch2}%
\end{align}
According to the previous output sample, they correspond to the resonances \#1 and \#2.

\subsection{All the semigroups with resonanes and/or zero element}
\label{gen_checkings_all}

In some applications it is useful to have a list containing all the resonant decompositions of semigroups of a given order, e.g., if we want to study all the possible resonant subalgebras of the S-expansions of a given Lie algebra. 
The program 13 perform this task for the order 4. 

\begin{small}
\begin{lstlisting}[language=Java]
package examples;
import sexpansion.Semigroup;
import sexpansion.SetS;
public class II_findAllResonances_console_ord4 {
 public static void main(String[] args) {
  Semigroup [] list = Semigroup.loadFromFile("src/data/");
  SetS [][] Resonances;
  SetS S0, S1;
  int i, j, nResonances=0, nSemigroupWithResonance=0, TotalResonances=0;
  for (i=0; i < list.length ; i++) {
   if (list[i].order == 4 && list[i].isCommutative()) {
    Resonances = list[i].findAllResonances();
    if (Resonances != null) {
     nSemigroupWithResonance = nSemigroupWithResonance +1;
     TotalResonances = TotalResonances + Resonances.length;
     System.out.println("The semigroup #"+list[i].ID+" has "
		           +Resonances.length+" resonances");
     list[i].show();
     for (j = 0; j < Resonances.length ; j++) {
      nResonances = nResonances + 1;
      S0 = Resonances[j][0];
      S1 = Resonances[j][1];
      System.out.println("Resonance #" +nResonances);
      System.out.print("S0: "); 
      S0.show();
      System.out.print("S1 :"); 
      S1.show();
     }
     nResonances = 0;
  }}}
  System.out.println("There are "+nSemigroupWithResonance+
   " semigroups with at least one resonance and there are"
    +" in total "+TotalResonances+" different resonances.");
}}
\end{lstlisting}
\end{small}
For space reasons we cannot give the full output here, but a sample looks as follows.
\begin{small}
\begin{verbatim}
...
The semigroup #42 has 5 resonances 
1 1 1 1  
1 1 1 2  
1 1 1 3  
1 2 3 4  
Resonance #1
S0: 1 4 
S1: 1 2 3 
Resonance #2
S0: 1 3 4 
S1: 1 2 
Resonance #3
S0: 1 2 4 
S1: 1 3 
Resonance #4
S0: 1 3 4 
S1: 1 2 3 
Resonance #5
S0: 1 2 4 
S1: 1 2 3 
The semigroup #43 has 2 resonances 
...
There are 48 semigroups with at least one resonance and 
there are in total 124 different resonances.
\end{verbatim}
\end{small}
Thus, apart from giving explicitly the resonances of all the non-isomorphic semigroups of a given order, the program 13 also gives the total number of semigroups having at least one resonance and the total number of different resonances. By changing the value of the variable \textit{list[i].order} one can easily obtain the results for other orders.
For the order 5 and 6 the output is so big that, depending on the Java environment that is being used, it could not be fully printed to screen. 
For this reason, the program 14 makes the same calculation for the orders $n=2,\ldots,6$ and prints the output in the folder\footnote{In particular, these outputs can also be found as \textit{pdf} documents in the file \textit{Output examples.zip} which is available together with the library in \cite{webJava}. Using font-size 10, those documents contain 13 pages for $n=4$, 130 for $n=5$ and 1668 for $n=6$.} \textit{``./Output\_examples/''}. 
\bigskip

On the other hand, also a list with all the semigroups of a given order having a zero element is useful if we want to study all the
possible S-expansions followed by a $0_{S}$-reduction for a given Lie algebra. 
The program 18 make this for the order 4.
\begin{small}
\begin{lstlisting}[language=Java]
package examples;
import sexpansion.Semigroup;
public class II_findzero_console_ord4 {
 public static void main(String[] args) {
  Semigroup[] list = Semigroup.loadFromFile("src/data/");
  Semigroup s ;
  int elementoCero ;
  int i,j;
  j = 0;
  for ( i = 0 ; i < list.length ; ++i){
   s = list[i];
   elementoCero = list[i].findZero();
   if ( s.order == 4 && elementoCero != -1 && list[i].isCommutative()) {
    ++j;
    System.out.println("#" +list[i].ID);
    list[i].show();
    System.out.print("The zero element is");
    System.out.println(elementoCero);
  }}
  System.out.println("Number of semigroups with zero element:" +j);
}}
\end{lstlisting}
\end{small}
Again, changing the value in the variable \textit{s.order} one can easily obtain the results for other orders. In particular, the program 20 performs the same calculation for $n=2,\ldots,6$ and prints the results in the folder \textit{``./Output\_examples/''}. 

Finally, the program 17 mix both types of codes described above in order to give the list of all semigroups having simultaneously both zero element and at least one resonance (the outputs are also printed to a file).
It is also worth to point out that the type of programs described here were used in the Sections 4.1 and 4.2 of Ref. \cite{Andrianopoli:2013ooa} to study the general properties of the expansions with semigroups up to order 6. These results will be summarized in Eq. \ref{cp_table1} of the Section \ref{pres_semi}, where also the number of semigroups preserving semisimplicity will be given.

\subsection{Examples of isomorphisms}
\label{gen_checkings2}

As shown in Refs. \cite{irs,Izaurieta:2009hz,Salgado:2014qqa}, the following families of semigroups %

\begin{align}
S_{E}^{\left(  N\right)  }  & =\left\{  \lambda_{0},\ldots,\lambda_{N}%
,\lambda_{N+1}\right\}  \text{ and }\lambda_{\alpha}\lambda_{\beta}=\left\{
\begin{tabular}
[c]{ll}%
$\lambda_{\alpha+\beta}\,,$ & $\alpha+\beta\leq N+1$\\
$\lambda_{N+1}\,,$ & $\alpha+\beta>N+1$%
\end{tabular}
\right.  \,,\label{SE}\\
S_{M}^{\left(  N\right)  }  & =\left\{  \lambda_{0},\ldots,\lambda
_{N}\right\}  \text{ and }\lambda_{\alpha}\lambda_{\beta}=\left\{
\begin{tabular}
[c]{ll}%
$\lambda_{\alpha+\beta}\,,$ & $\alpha+\beta\leq6$\\
$\lambda_{\alpha+\beta-6}\,,$ & $\alpha+\beta>6$%
\end{tabular}
\right.  \,,\label{SM}%
\end{align}
allows to generate the algebras $\mathfrak{B}_{N+2}$ and $\mathfrak{C}_{N+2}$
which, as explained in the introduction, have been recently used in many applications in gravity theories. 
For each order $n$, the semigroups $S_{E}^{\left(  n-2\right)  }$
and $S_{M}^{\left(  n-1\right)  }$ represents non isomorphic semigroups. For example, the semigroups corresponding of order $7$ are given by,
\begin{equation}%
\begin{tabular}
[c]{l|lllllll}%
$S_{E}^{\left(  5\right)  }$ & $\lambda_{0}$ & $\lambda_{1}$ & $\lambda_{2}$ &
$\lambda_{3}$ & $\lambda_{4}$ & $\lambda_{5}$ & $\lambda_{6}$\\\hline
$\lambda_{0}$ & $\lambda_{0}$ & $\lambda_{1}$ & $\lambda_{2}$ & $\lambda_{3}$
& $\lambda_{4}$ & $\lambda_{5}$ & $\lambda_{6}$\\
$\lambda_{1}$ & $\lambda_{1}$ & $\lambda_{2}$ & $\lambda_{3}$ & $\lambda_{4}$
& $\lambda_{5}$ & $\lambda_{6}$ & $\lambda_{6}$\\
$\lambda_{2}$ & $\lambda_{2}$ & $\lambda_{3}$ & $\lambda_{4}$ & $\lambda_{5}$
& $\lambda_{6}$ & $\lambda_{6}$ & $\lambda_{6}$\\
$\lambda_{3}$ & $\lambda_{3}$ & $\lambda_{4}$ & $\lambda_{5}$ & $\lambda_{6}$
& $\lambda_{6}$ & $\lambda_{6}$ & $\lambda_{6}$\\
$\lambda_{4}$ & $\lambda_{4}$ & $\lambda_{5}$ & $\lambda_{6}$ & $\lambda_{6}$
& $\lambda_{6}$ & $\lambda_{6}$ & $\lambda_{6}$\\
$\lambda_{5}$ & $\lambda_{5}$ & $\lambda_{6}$ & $\lambda_{6}$ & $\lambda_{6}$
& $\lambda_{6}$ & $\lambda_{6}$ & $\lambda_{6}$\\
$\lambda_{6}$ & $\lambda_{6}$ & $\lambda_{6}$ & $\lambda_{6}$ & $\lambda_{6}$
& $\lambda_{6}$ & $\lambda_{6}$ & $\lambda_{6}$%
\end{tabular}
\,,\ \
\begin{tabular}
[c]{l|lllllll}%
$S_{M}^{\left(  6\right)  }$ & $\lambda_{0}$ & $\lambda_{1}$ & $\lambda_{2}$ &
$\lambda_{3}$ & $\lambda_{4}$ & $\lambda_{5}$ & $\lambda_{6}$\\\hline
$\lambda_{0}$ & $\lambda_{0}$ & $\lambda_{1}$ & $\lambda_{2}$ & $\lambda_{3}$
& $\lambda_{4}$ & $\lambda_{5}$ & $\lambda_{6}$\\
$\lambda_{1}$ & $\lambda_{1}$ & $\lambda_{2}$ & $\lambda_{3}$ & $\lambda_{4}$
& $\lambda_{5}$ & $\lambda_{6}$ & $\lambda_{1}$\\
$\lambda_{2}$ & $\lambda_{2}$ & $\lambda_{3}$ & $\lambda_{4}$ & $\lambda_{5}$
& $\lambda_{6}$ & $\lambda_{1}$ & $\lambda_{2}$\\
$\lambda_{3}$ & $\lambda_{3}$ & $\lambda_{4}$ & $\lambda_{5}$ & $\lambda_{6}$
& $\lambda_{1}$ & $\lambda_{2}$ & $\lambda_{3}$\\
$\lambda_{4}$ & $\lambda_{4}$ & $\lambda_{5}$ & $\lambda_{6}$ & $\lambda_{1}$
& $\lambda_{2}$ & $\lambda_{3}$ & $\lambda_{4}$\\
$\lambda_{5}$ & $\lambda_{5}$ & $\lambda_{6}$ & $\lambda_{1}$ & $\lambda_{2}$
& $\lambda_{3}$ & $\lambda_{4}$ & $\lambda_{5}$\\
$\lambda_{6}$ & $\lambda_{6}$ & $\lambda_{1}$ & $\lambda_{2}$ & $\lambda_{3}$
& $\lambda_{4}$ & $\lambda_{5}$ & $\lambda_{6}$%
\end{tabular}
\label{Ex1_Isom}%
\end{equation}
The program 4, 
\begin{small}
\begin{lstlisting}[language=Java]
package examples;
import sexpansion.Semigroup;
public class I_isomorphisms_ex1 {
 public static void main(String[] args) {
\end{lstlisting}
\end{small}
determines if there are isomorphisms between them. After introducing the semigroups as done before (reminding to rename the semigroups elements in such a way that they are represented by positive interger numbers), we use the method \textit{isotest} as follows,
\begin{small}
\begin{lstlisting}[language=Java]
  if ( SE_5.isotest(SM_6)[0]) {
   System.out.println("SE_5 is isomorphic to SM_6");
  }
  else {
   System.out.println("SE_5 is not isomorphic to SM_6");
}}}
\end{lstlisting}
\end{small}
The output reads,
\begin{small}
\begin{verbatim}
SE_5 is not isomorphic to SM_6
\end{verbatim}
\end{small}

Similar programs were used in Ref. \cite{Caroca:2011qs} to check that the semigroups $S_{E}^{(2)}$, $S_{K}^{(3)}$, $S_{N1}$, $S_{N2}$, $S_{N3}$ are not isomorphic between them and thus, that they represent different solutions to relate 2 and 3-dimensional algebras. Indeed, with the same kind of code, one could easily check that all the semigroups of order 4 given in Eq. (\ref{Ex_B_CS}) are not isomorphic between them.
\bigskip

On the other hand, there are cases where we need to identify the specific semigroup $S_{(n)}^{a}$ of the lexicographical classification\footnote{We remind that those semigroups were labeled in Section \ref{hist_semig} by $S_{(n)}^{a}$, where $a=1,...,Q$ identify a specific semigroup of order $n$ and $Q$ is the total number of semigroups in that order.} to which a given semigroup $\tilde{S}$ is isomorphic. For example, let us consider the semigroup $S_{N3}$ of the Eq. (\ref{Ex_B_CS}). Using the method \textit{isEqualTo} of the \textit{Semigroup} class one may check that its multiplication table does not coincide with any table of the list \textit{sem.4} (and, in general, this will happens for any semigroup table which does not have a lexicographical order). However that list is exhaustive, i.e., it contains all the non isomorphic semigroups of order 4 and thus, $S_{N3}$ must be isomorphic to one and only one semigroup in the list \textit{sem.4}. 

Before showing how we can find the semigroup $S_{(4)}^{a}$ which is isomorphic to $S_{N3}$, it is worth to see first how the methods of our library generate all the permutations (and the corresponding inverses) of a given set with $n$ elements. The program 8 perform that task for $n=4$ using the methods \textit{allPermutations} and \textit{inversePermutation}. First, it defines a \textit{SetS} object \textit{n\_elements} with $n$ elements and stores all its permutations in the \textit{SetS} object called \textit{allpermut}. 
\begin{small}
\begin{lstlisting}[language=Java]
package examples;
import sexpansion.SetS;
public class I_Permutations_and_inverses_console_n4 {
 public static void main(String[] args) {
  int n = 4, i;
  SetS [] allpermut ;
  SetS n_elements ;
  n_elements = new SetS(n);
  allpermut = n_elements.allPermutations() ;
\end{lstlisting}
\end{small}
Then, with the following piece of code we obtain the result.
\begin{small}
\begin{lstlisting}[language=Java]
  for ( i = 0 ; i < allpermut.length; ++i ) {
   System.out.println("Permutation #" +i);
   allpermut[i].show();
   System.out.println("The inverse permutation is:");
   allpermut[i].inversePermutation().show();
}}}
\end{lstlisting}
\end{small}
In the same way, the program 9 generate the list of all permutations with their corresponding inverses for a set of $n$ elements, with $n=2,\ldots,7$. The results are direcly printed in the folder \textit{``./Output\_examples/''} and in each case the $n!$ permutations are labeled here by $P^{\#X}$, with $X=0,...,(n!-1)$.

Now, it is easy to see how the program 5 finds the semigroup $S_{(4)}^{a}$ of the lexicographic classification which is isomorphic to $S_{N3}$.
First, it reads the semigroup $S_{N3}$, the full list of semigroups of order 4 and defines some useful variables with the method \textit{allPermutations} explained before.
\begin{small}
\begin{lstlisting}[language=Java]
package examples;
import sexpansion.Semigroup;
import sexpansion.SetS;
public class I_isomorphisms_ex2 {
 public static void main(String[] args) {
  int [][] mSN3 = {{4,1,4,4},{1,2,3,4},{4,3,4,4},{4,4,4,4}};
  Semigroup SN3 = new Semigroup(mSN3);
  Semigroup[] lista = Semigroup.loadFromFile("src/data/");
  int i,k ;
  Semigroup [] permutations ;
  SetS [] p = (new SetS(4)).allPermutations();
\end{lstlisting}
\end{small}
After introducing $S_{N3}$ as usual, the following piece of code performs the task.
\begin{small}
\begin{lstlisting}[language=Java]
  for ( i = 0 ; i < lista.length ; ++i){
   if ( lista[i].order == 4  ) {
    if ( lista[i].isotest(SN3)[0]) {
     System.out.println("The semigroup #" + lista[i].ID);
     lista[i].show();
     System.out.println("is isomorphic to SN3.");
     permutations = lista[i].permute();
     for ( k = 0 ; k < permutations.length; ++k) {
      if ( permutations[k].isEqualTo(SN3)) {
        System.out.print( "A permutation that brings #" 
      + lista[i].ID + " to SN3 is P#");
        System.out.println(k);
        p[k].show();
        System.out.println("The inverse permutation is:");
        p[k].inversePermutation().show();
}}}}}}}
\end{lstlisting}
\end{small}
The output reads,
\begin{small}
\begin{verbatim}
The semigroup #42
1 1 1 1  
1 1 1 2  
1 1 1 3  
1 2 3 4  
is isomorphic to SN3.
****
A permutation that brings #42 to SN3 is P#19
4 1 3 2 
The inverse permutation is:
2 4 3 1 
----
A permutation that brings #42 to SN3 is P#22
4 3 1 2 
The inverse permutation is:
3 4 2 1 
\end{verbatim}
\end{small}
Interestingly, more than one permutation may represent the isomorphism. To show how the permutation must be applied according to the definition given by Eq. (\ref{perm_not}), let us see explicitly the relation with $P^{\#19}\,$:
\begin{equation}
P^{\#19}\left(
\begin{tabular}
[c]{l|llll}%
$S_{\left(  4\right)  }^{42}$ & $\lambda_{1}$ & $\lambda_{2}$ & $\lambda_{3}$
& $\lambda_{4}$\\\hline
$\lambda_{1}$ & $\lambda_{1}$ & $\lambda_{1}$ & $\lambda_{1}$ & $\lambda_{1}%
$\\
$\lambda_{2}$ & $\lambda_{1}$ & $\lambda_{1}$ & $\lambda_{1}$ & $\lambda_{2}%
$\\
$\lambda_{3}$ & $\lambda_{1}$ & $\lambda_{1}$ & $\lambda_{4}$ & $\lambda_{3}%
$\\
$\lambda_{4}$ & $\lambda_{1}$ & $\lambda_{2}$ & $\lambda_{3}$ & $\lambda_{4}$%
\end{tabular}
\right)  =%
\begin{tabular}
[c]{l|llll}
& $\lambda_{4}$ & $\lambda_{1}$ & $\lambda_{3}$ & $\lambda_{2}$\\\hline
$\lambda_{4}$ & $\lambda_{4}$ & $\lambda_{4}$ & $\lambda_{4}$ & $\lambda_{4}%
$\\
$\lambda_{1}$ & $\lambda_{4}$ & $\lambda_{4}$ & $\lambda_{4}$ & $\lambda_{1}%
$\\
$\lambda_{3}$ & $\lambda_{4}$ & $\lambda_{4}$ & $\lambda_{4}$ & $\lambda_{3}%
$\\
$\lambda_{2}$ & $\lambda_{4}$ & $\lambda_{1}$ & $\lambda_{3}$ & $\lambda_{2}$%
\end{tabular}
\sim%
\begin{tabular}
[c]{l|llll}%
$S_{N3}$ & $\lambda_{1}$ & $\lambda_{2}$ & $\lambda_{3}$ & $\lambda_{4}%
$\\\hline
$\lambda_{1}$ & $\lambda_{4}$ & $\lambda_{1}$ & $\lambda_{4}$ & $\lambda_{4}%
$\\
$\lambda_{2}$ & $\lambda_{1}$ & $\lambda_{2}$ & $\lambda_{3}$ & $\lambda_{4}%
$\\
$\lambda_{3}$ & $\lambda_{4}$ & $\lambda_{3}$ & $\lambda_{4}$ & $\lambda_{4}%
$\\
$\lambda_{4}$ & $\lambda_{4}$ & $\lambda_{4}$ & $\lambda_{4}$ & $\lambda_{4}$%
\end{tabular}
\,.\label{P19}%
\end{equation}

In addition, the program 10 uses the methods \textit{permuteWith} and \textit{equalTo} to check the result in both senses, i.e, it checks that:
\begin{itemize}
	\item $S_{N3}$ can be obtained by applying the permutations $P^{\#19}=(4\, 1\, 3\, 2)$  and $P^{\#22}=(4\, 3\, 1\, 2)$ to the semigroup $S_{(4)}^{42}$,
	\item $S_{(4)}^{42}$ can be obtained by applying the permutations the permutations $(P^{\#19})^{-1} = P^{\#11}=(2\, 4\, 3\, 1)$ and $(P^{\#22})^{-1} = P^{\#17}=(3\, 4\, 2\, 1)$ to the semigroup $S_{N3}$.
\end{itemize}
With those isomorphisms one can directly check that the five resonances found for $S_{N3}$ and $S_{(4)}^{42}$, respectively in Sections \ref{gen_checkings} and \ref{gen_checkings_all}, are in one to one correspondence.

With simple modifications on the inputs of the program described above, the reader may check if there exist isomorphisms for any given set of semigroups. Indeed, this type of code was used to find the isomorphisms given in Section 6 of Ref. \cite{Caroca:2011qs}.

\subsection{Examples with S-expanded algebras}
\label{gen_checkings3}

As explained in Section \ref{Rep_S_G}, given a semigroup we can use the method \textit{getSelector} to obtain the adjoint representation of any semigroup. An explicit example to do this is provided by the program 26.
Now, we are going to describe the main characteristics of the programs 28-35 in order to ilustrate how to perform S-expansions.

Let us first consider the Lie algebra $\mathcal{G}=\mathfrak{sl}(2)$ with the following commutation relations
\begin{align}
&  \left[  X_{1},X_{2}\right]  =-2X_{3}\nonumber\\
&  \left[  X_{1},X_{3}\right]  =2X_{2}\nonumber\\
&  \left[  X_{2},X_{3}\right]  =2X_{1}\,.\label{sl2_com2}%
\end{align}
Consider also the semigroup
\begin{equation}%
\begin{tabular}
[c]{l|lllll}%
$S_{\left(  5\right)  }^{770}$ & $\lambda_{1}$ & $\lambda_{2}$ & $\lambda_{3}$
& $\lambda_{4}$ & $\lambda_{5}$\\\hline
$\lambda_{1}$ & $\lambda_{1}$ & $\lambda_{1}$ & $\lambda_{1}$ & $\lambda_{1}$
& $\lambda_{1}$\\
$\lambda_{2}$ & $\lambda_{1}$ & $\lambda_{2}$ & $\lambda_{1}$ & $\lambda_{1}$
& $\lambda_{5}$\\
$\lambda_{3}$ & $\lambda_{1}$ & $\lambda_{1}$ & $\lambda_{3}$ & $\lambda_{4}$
& $\lambda_{1}$\\
$\lambda_{4}$ & $\lambda_{1}$ & $\lambda_{1}$ & $\lambda_{4}$ & $\lambda_{3}$
& $\lambda_{1}$\\
$\lambda_{5}$ & $\lambda_{1}$ & $\lambda_{5}$ & $\lambda_{1}$ & $\lambda_{1}$
& $\lambda_{2}$%
\end{tabular}
\,,\label{S770}%
\end{equation}
which has the resonant decomposition $S_{0}=\left\{  \lambda_{1},\lambda_{2},\lambda_{3}\right\}$, $S_{0}=\left\{\lambda_{1},\lambda_{4},\lambda_{5}\right\}$ and where $\lambda_{1}$ is the zero element. 
The program 28 gives the non vanishing commutators and structure constants of:
\begin{itemize}
	\item the expanded algebra $\mathcal{G}_{S}$,
	\item the resonant subalgebra $\mathcal{G}_{S,R}$,
	\item the reduced algebra $\mathcal{G}_{S,\text{red}}$ and
	\item the reduction of the resonant subalgebra $\mathcal{G}_{S,R,\text{red}}$
\end{itemize}
To do this we need the following classes.
\begin{small}
\begin{lstlisting}[language=Java]
package examples;
import Jama.Matrix;
import sexpansion.Semigroup;
import sexpansion.SetS;
import sexpansion.StructureConstantSet;
import sexpansion.StructureConstantSetExpanded;
import sexpansion.StructureConstantSetExpandedReduced;
import sexpansion.StructureConstantSetExpandedResonant;
import sexpansion.StructureConstantSetExpandedResonantReduced;
public class II_SExp_sl2_S770 {
 public static void main(String[] args) {
\end{lstlisting}
\end{small}

Then, we define an object called \textit{metric} with the class \textit{Matrix} of the \textit{jama} library and use the method \textit{setStructureConstant} of the class \textit{StructureConstantSet} to introduce\footnote{According to the convention explained at the end of Section \ref{Rep_S_G}, this is done in such a way that $i,j,k=0,1,\ldots,n-1$ where $n$ is the dimension of $\mathcal{G}$. Similarly $a,b,c=0,1,\ldots,m-1$ in the functions $C_{(i,a)(j,b)}^{(k,c)}$, where $m$ is the order of the semigroup. However, the outputs will be given in such a way that $i,j,k=1,\ldots,n$ and $a,b,c=1,\ldots,m$.} the non vanishing structure constants $C_{ij}^{k}$ of the original algebra.
\begin{small}
\begin{lstlisting}[language=Java]
  Matrix metric ;
  StructureConstantSet sl2 = new StructureConstantSet(3) ;
  sl2.setStructureConstant(0, 1, 2,-2) ;
  sl2.setStructureConstant(0, 2, 1, 2) ;
  sl2.setStructureConstant(1, 2, 0, 2) ;
  metric = sl2.cartanKillingMetric() ;
  metric.print(2,2); // Show its KC metric
  System.out.println(metric.det()); // Prints its determinant
\end{lstlisting}
\end{small}
In addition the KC metric and its determinant is calculated.
Next, the program loads the semigroup $S_{(5)}^{770}$ and its resonant decomposition using the classes \textit{Semigroup} and \textit{SetS}
\begin{small}
\begin{lstlisting}[language=Java]
  int [][] mS770 = {{1,1,1,1,1},{1,2,1,1,5},{1,1,3,4,1},
                   {1,1,4,3,1},{1,5,1,1,2}};
  Semigroup S770 = new Semigroup(mS770) ;
  int[] mS0 = {1,2,3} ;
  SetS S0 = new SetS(mS0) ;
  int[] mS1 = {1,4,5} ;
  SetS S1 = new SetS(mS1) ;
  int[] mV0 = {1};
  SetS V0 = new SetS(mV0);
  int[] mV1 = {2,3};
  SetS V1 = new SetS(mV1); 
\end{lstlisting}
\end{small}
Now, we use the methods \textit{showCommut}, \textit{showSC} and \textit{cartanKillingMetric} to compute the commutators, structure constants and KC metric of expanded algebra
(here, some prints which explain the output are omitted for space reasons).
\begin{small}
\begin{lstlisting}[language=Java]
  // 1) the expanded algebra
  StructureConstantSetExpanded Gexp = S770.getExpandedStructureConstant(sl2);
  Gexp.showCommut();
  Gexp.showSC();
  Matrix metricaGexp = Gexp.cartanKillingMetric();
  metricaGexp.print(0,0);
  System.out.println(metricaGexp.det());
\end{lstlisting}
\end{small}
This way, we have reproduced the 6 steps explained in Section \ref{Rep_GS} for the specific expansion of $\mathfrak{sl}(2)$ with the semigroup $S_{\left(5\right)}^{770}$.

Similarly, the program 28 follows the steps given in Sections \ref{Rep_GSRes} and \ref{Rep_GSRed} to obtain the resonant subalgebra and reduced algebra. 
\begin{small}
\begin{lstlisting}[language=Java]
  // 2) the resonant subalgebra
  StructureConstantSetExpandedResonant ResGexp = 
   new StructureConstantSetExpandedResonant(Gexp.data,S0,S1,V0,V1);
  ResGexp.showCommutRes();
  ResGexp.showSCRes();
  Matrix metricaResGexp = ResGexp.cartanKillingMetricPretty();
  metricaResGexp.print(0,0);
  System.out.println(metricaResGexp.det());
  // 3) the reduced algebra showCommutRel 
  StructureConstantSetExpandedReduced RedGexp = 
   new StructureConstantSetExpandedReduced
   (S770.getExpandedStructureConstant(sl2).data,1);
  RedGexp.showCommutRed();
  RedGexp.showSCRed();
  Matrix metricaRedGexp = RedGexp.cartanKillingMetricPretty();
  metricaRedGexp.print(0,0);
  System.out.println(metricaRedGexp.det());
\end{lstlisting}
\end{small}

Finally, the reduction of the resonant subalgebra is obtained as follows.
\begin{small}
\begin{lstlisting}[language=Java]
  // 4) the reduction of the resonant subalgebra  
  StructureConstantSetExpandedResonantReduced RedResGexp = 
   new StructureConstantSetExpandedResonantReduced(ResGexp,1) ;
  RedResGexp.showCommutResRed();
  RedResGexp.showSCResRed();
  Matrix metricaRedResGexp = RedResGexp.cartanKillingMetricPretty();
  metricaRedResGexp.print(0,0);
  System.out.println(metricaRedResGexp.det());
}}
\end{lstlisting}
\end{small}

A sample output for the expanded algebra looks as follows.
\begin{small}
\begin{verbatim}
Killing-Cartan metric of sl(2)
 -8.00 0.00 0.00
  0.00 8.00 0.00
  0.00 0.00 8.00
whose determinant is: -512.0

METHOD: showCommut()
Non vanishing commutators of the 'Expanded algebra'

n = 3 , Dimension of the original Lie algebra.
m = 5 , Order of the semigroup.

With the notation: X_{i,a}= X_{i} lambda_{a}, the generators of the 
'Expanded algebra' are given by:
 Y_{1}  =  X_{1,1}
 Y_{2}  =  X_{1,2}
 Y_{3}  =  X_{1,3}
 Y_{4}  =  X_{1,4}
 Y_{5}  =  X_{1,5}
 Y_{6}  =  X_{2,1}
 Y_{7}  =  X_{2,2}
 Y_{8}  =  X_{2,3}
 Y_{9}  =  X_{2,4}
 Y_{10} =  X_{2,5}
 Y_{11} =  X_{3,1}
 Y_{12} =  X_{3,2}
 Y_{13} =  X_{3,3}
 Y_{14} =  X_{3,4}
 Y_{15} =  X_{3,5}

The non vanishing commutators of the 'Expanded algebra' are given by:
 [ X_{1,1} , X_{2,1} ] = -2.0 X_{3,1}
 [ X_{1,1} , X_{2,2} ] = -2.0 X_{3,1}
... // here it follow 73 non vanishing commutators of the Expanded algebra

METHOD: showSC()
Non vanishing structure constants of the Expanded algebra:
 C_{(1,1)(2,1)}^{(3,1)} = -2.0
 C_{(1,1)(2,2)}^{(3,1)} = -2.0
... // here it follow 73 non vanishing structure constants of the Expanded algebra 

METHOD: cartanKillingMetric()
The Killing-Cartan Metric of the Expanded algebra is:

 -8  -8  -8  -8  -8  0  0  0  0  0  0  0  0  0  0
 -8 -24  -8  -8  -8  0  0  0  0  0  0  0  0  0  0
 -8  -8 -24  -8  -8  0  0  0  0  0  0  0  0  0  0
 -8  -8  -8 -24  -8  0  0  0  0  0  0  0  0  0  0
 -8  -8  -8  -8 -24  0  0  0  0  0  0  0  0  0  0
  0   0   0   0   0  8  8  8  8  8  0  0  0  0  0
  0   0   0   0   0  8 24  8  8  8  0  0  0  0  0
  0   0   0   0   0  8  8 24  8  8  0  0  0  0  0
  0   0   0   0   0  8  8  8 24  8  0  0  0  0  0
  0   0   0   0   0  8  8  8  8 24  0  0  0  0  0
  0   0   0   0   0  0  0  0  0  0  8  8  8  8  8
  0   0   0   0   0  0  0  0  0  0  8 24  8  8  8
  0   0   0   0   0  0  0  0  0  0  8  8 24  8  8
  0   0   0   0   0  0  0  0  0  0  8  8  8 24  8
  0   0   0   0   0  0  0  0  0  0  8  8  8  8 24

The determinant of the 
Killing-Cartan Metric of the Expanded algebra is:
-1.44115188075855872E17
\end{verbatim}
\end{small}
Now, the part of the output corresponding to the reduction of the resonant sublgebra looks as follows.
\begin{small}
\begin{verbatim}
METHOD: showCommutResRed()
Non vanishing commutators of the 'Reduction of the Resonant Subalgebra'

With the notation: X_{i,a}= X_{i} lambda_{a}, the generators of the 
'Reduction of the Resonant Subalgebra' are given by:
 Y_{2}  =  X_{1,2}
 Y_{3}  =  X_{1,3}
 Y_{9}  =  X_{2,4}
 Y_{10} =  X_{2,5}
 Y_{14} =  X_{3,4}
 Y_{15} =  X_{3,5}

The non vanishing commutators of the 'Reduction of the Resonant Subalgebra' 
are given by:
 [ X_{1,2} , X_{2,5} ] = -2.0 X_{3,5}
 [ X_{1,2} , X_{3,5} ] =  2.0 X_{2,5}
 [ X_{1,3} , X_{2,4} ] = -2.0 X_{3,4}
 [ X_{1,3} , X_{3,4} ] =  2.0 X_{2,4}
 [ X_{2,4} , X_{3,4} ] =  2.0 X_{1,3}
 [ X_{2,5} , X_{3,5} ] =  2.0 X_{1,2}

METHOD: showSCResRed()
Non vanishing structure constants of the 'Reduction of the Resonant Subalgebra' 
are given by:
 C_{(1,2)(2,5)}^{(3,5)} = -2.0
 C_{(1,2)(3,5)}^{(2,5)} =  2.0
 C_{(1,3)(2,4)}^{(3,4)} = -2.0
 C_{(1,3)(3,4)}^{(2,4)} =  2.0
 C_{(2,4)(3,4)}^{(1,3)} =  2.0
 C_{(2,5)(3,5)}^{(1,2)} =  2.0

METHOD: cartanKillingMetricPretty()
The Killing-Cartan Metric of the Reduced algebra is:

 -8  0  0  0  0  0
  0 -8  0  0  0  0
  0  0  8  0  0  0
  0  0  0  8  0  0
  0  0  0  0  8  0
  0  0  0  0  0  8

The determinant of the Killing-Cartan Metric of the Reduced algebra is:
262144.0
-------------------
\end{verbatim}
\end{small}

The programs 29-31 perform the same kind of calculation for the semigroups $S_{\left(5\right)}^{968}$, $S_{\left(5\right)}^{990}$ and $S_{\left(5\right)}^{991}$. 
According to the Section 5 of Ref. \cite{Andrianopoli:2013ooa} these are the semigroups of lowest order that leads (after extracting reduced and resonant subalgebras) to non trivial expansions of $\mathfrak{sl}(2)$ in such a way that semisimplicity is preserved. 
In addition the programs 32-35 uses the method \textit{showPretty} to generate the adjoint representations for expansions of $\mathfrak{sl}(2)$ with the same semigroups. A sample output for the reduction of the resonant subalgebra obtained with the semigroup $S_{\left(5\right)}^{770}$ is given in what follows.
\begin{small}
\begin{verbatim}
NOTATION for the Reduction of the Resonant Subalgebra:

To print the structure constants notice that for (i,a) fixed,
the quantities C_{(i,a)(j,b)}^{(k,c)}=M_{A,B} are elements
of a matrix M whose indices have the following values:
A,B = 2, 3, 9, 10, 14, 15, 
Or equivalently, 
A,B = (1,2), (1,3), (2,4), (2,5), (3,4), (3,5), 

Here we print the matrices C_{(1,a) (j,b)}^{(k,c)}, with the double indices 
having the values described above.
******
C_{(1,2) (j,b)}^{(k,c)}
 0.0  0.0  0.0  0.0  0.0  0.0 
 0.0  0.0  0.0  0.0  0.0  0.0 
 0.0  0.0  0.0  0.0 -0.0 -0.0 
 0.0  0.0  0.0  0.0 -0.0 -2.0 
 0.0  0.0  0.0  0.0  0.0  0.0 
 0.0  0.0  0.0  2.0  0.0  0.0 
******
C_{(1,3) (j,b)}^{(k,c)}
 0.0  0.0  0.0  0.0  0.0  0.0 
 0.0  0.0  0.0  0.0  0.0  0.0 
 0.0  0.0  0.0  0.0 -2.0 -0.0 
 0.0  0.0  0.0  0.0 -0.0 -0.0 
 0.0  0.0  2.0  0.0  0.0  0.0 
 0.0  0.0  0.0  0.0  0.0  0.0 
*****
Here we print the matrices C_{(2,a) (j,b)}^{(k,c)}, with the double indices 
having the values described above.
...
\end{verbatim}
\end{small}

The full output of the programs described on this section can be found in the file \textit{Output\_examples.zip} available in \cite{webJava}.

\subsection{Identifying S-expansions that preserve semisimplicity}
\label{pres_semi}

In \cite{Andrianopoli:2013ooa} it was shown that commutativity, solvability and nilpotency are properties preserved under the S-expansions any semigroup, whilst properties like semisimplicity and compactness are preserved only for semigroups satisfying certain conditions (see the summary given in Section \ref{sexpp}).
In what follows, we will show how our library can be used to identify all the semigroups preserving semisimplicity up to order 6.

Let us consider the $\mathfrak{sl}(2)$ algebra, whose commutation relations are given in Eq. (\ref{sl2_com2}). The program 42 calculates 
the S-expanded algebras $\mathcal{G}_{S}= S_{(n)}^{a} \otimes\mathcal{G}$ for all the non-isomorphic semigroups of the lists \textit{sem.n} for $n=2, \cdots ,6$.
As shown in what follows, for each expansion it checks if the determinant of the expanded Killing-Cartan metric is different than zero (again, for space reasons, some prints explaining the output are omitted)
\begin{small}
\begin{lstlisting}[language=Java]
package examples;
import Jama.Matrix;
import sexpansion.Semigroup;
import sexpansion.StructureConstantSet;
import sexpansion.StructureConstantSetExpanded;
public class III_sl2_SExp_ord2_to_6_console {
// To change the order modify the value of the interger 'order'
// (only for values between 2 and 6)
 private static int order=3 ;
 public static void main(String[] args) {
  Matrix metric ;
  Semigroup [] listSemigroups = Semigroup.loadFromFile("src/data/") ;
  StructureConstantSet sl2 = new StructureConstantSet(3) ;
  sl2.setStructureConstant(0, 1, 2, -2) ;
  sl2.setStructureConstant(0, 2, 1, 2) ;
  sl2.setStructureConstant(1, 2, 0, 2) ;
  metric = sl2.cartanKillingMetric() ;
  metric.print(2, 2) ;
  System.out.println(metric.det());
  Semigroup semigroup ;
  StructureConstantSetExpanded expandedAlgebra ;
  int nCommutativos = 0 ;
  int nSemisimples = 0 ;
  int i ;
  for ( i = 0 ; i < listSemigroups.length ; ++i) {
   semigroup = listSemigroups[i] ;
   if ( semigroup.order == order  && semigroup.isCommutative()) {
    nCommutativos = nCommutativos + 1 ;
    expandedAlgebra  = semigroup.getExpandedStructureConstant( sl2 );
    metric = expandedAlgebra.cartanKillingMetric() ;
    if (  metric.det() !=  0 ) {
     nSemisimples = nSemisimples + 1 ;
     System.out.println("A semisimple algebra has been found"
        +"expanding with the semigroup #"+semigroup.ID);
     semigroup.show();
     System.out.println("The metric of the expanded algebra is:");
     metric.print(2, 2);
     System.out.print("whose determinant is: ");
     System.out.println(metric.det());
  }}}
  System.out.print("There are "+nCommutativos+
     " commutative semigroups of order "+order+" and "+nSemisimples+
     " expansions that give a semisimple algebra.");
}}
\end{lstlisting}
\end{small}
A sample of output reads,
\begin{small}
\begin{verbatim}
Expanding by the semigroup #16
We have found a semisimple algebra
We have expanded by the semigroup:
1 1 3  
1 2 3  
3 3 1  
We show the metric of the S-expanded algebra
 16.00  16.00   0.00  0.00   0.00  0.00  0.00   0.00  0.00
 16.00  24.00   0.00  0.00   0.00  0.00  0.00   0.00  0.00
  0.00   0.00  16.00  0.00   0.00  0.00  0.00   0.00  0.00
  0.00   0.00   0.00  0.00   0.00  0.00  8.00   8.00  0.00
  0.00   0.00   0.00  0.00   0.00  0.00  8.00  12.00  0.00
  0.00   0.00   0.00  0.00   0.00  0.00  0.00   0.00  8.00
  0.00   0.00   0.00  8.00   8.00  0.00  0.00   0.00  0.00
  0.00   0.00   0.00  8.00  12.00  0.00  0.00   0.00  0.00
  0.00   0.00   0.00  0.00   0.00  8.00  0.00   0.00  0.00
Determinant of the metric: -1.34217728E8 .
\end{verbatim}
\end{small}

\begin{figure}[th]
\centering
\includegraphics[scale=0.6]{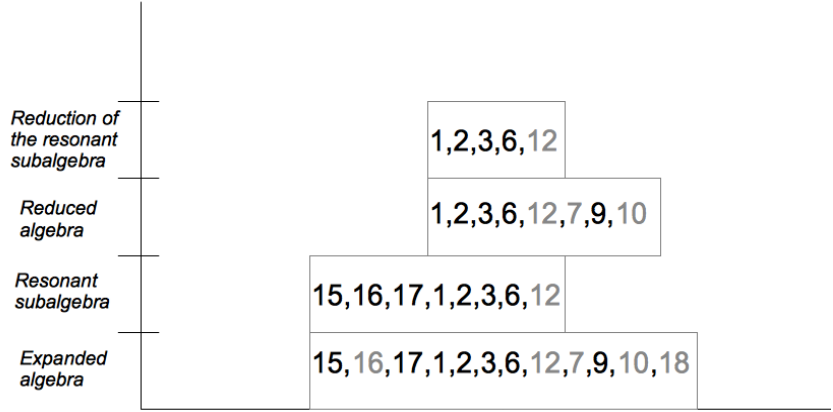}\caption{Expansions of
$\mathfrak{sl}(2,\mathbb{R})$ with abelian semigroups of order $3$}%
\label{fig:fig0}%
\end{figure}

Similarly, the programs 43-45 were used in Ref. \cite{Andrianopoli:2013ooa} to perform S-expansions with all the non-isomorphic semigroups which have zero element and/or at least one resonances. In each case, these programs dentify the expanded algebras $\mathcal{G}_{S,R}$, $\mathcal{G}_{S,\text{red}}$, $\mathcal{G}_{S,R,\text{red}}$ which are semisimple. 
In particular, the figure \ref{fig:fig0} ilustrate the different kind of expansions that can be done using the results of these programs for the order 3.
For space reasons, only the label `$a$' of the semigroup $S_{(3)}^{a}$ was written in the figure. The vertical axis illustrates the different types of expansions that can be performed
and the boxes over the horizontal axis contain the semigroups that can be used for each case. The semigroups that preserve semisimplicity are labeled with a \textit{gray} number and, as we can see, it might happen that a semigroup preserves semisimplicity in one level but not in other. This is the case of the semigroup $S_{(3)}^{16}$, appearing in the sample output above, which preserves semisimplicity for the expanded algebra but not for the resonant subalgebra. 

The results of the programs 42-45 (which can also be found in the file \textit{Output\_examples.zip} available in \cite{webJava}) are sumarized in the following table.
\begin{equation}%
\begin{tabular}
[c]{|l|c|c|c|c|c|}\hline
& $n=2$ & $n=3$ & $n=4$ & $n=5$ & $n=6$\\\hline%
\begin{tabular}
[c]{|l|}\hline
\#$\mathcal{G}_{S}$\\\hline
\#{\small pss}\\\hline
\end{tabular}
& \multicolumn{1}{|r|}{%
\begin{tabular}
[c]{|l|}\hline
$3$\\\hline
$2$\\\hline
\end{tabular}
} & \multicolumn{1}{|r|}{%
\begin{tabular}
[c]{|r|}\hline
$12$\\\hline
$5$\\\hline
\end{tabular}
} & \multicolumn{1}{|r|}{%
\begin{tabular}
[c]{|l|}\hline
$58$\\\hline
$16$\\\hline
\end{tabular}
} & \multicolumn{1}{|r|}{%
\begin{tabular}
[c]{|l|}\hline
$325$\\\hline
\multicolumn{1}{|r|}{$51$}\\\hline
\end{tabular}
} & \multicolumn{1}{|r|}{%
\begin{tabular}
[c]{|l|}\hline
$2,143$\\\hline
\multicolumn{1}{|r|}{$201$}\\\hline
\end{tabular}
}\\\hline%
\begin{tabular}
[c]{|l|}\hline
\#$\mathcal{G}_{S,\text{red}}$\\\hline
\#{\small pss}\\\hline
\end{tabular}
& \multicolumn{1}{|r|}{%
\begin{tabular}
[c]{|l|}\hline
$2$\\\hline
$1$\\\hline
\end{tabular}
} & \multicolumn{1}{|r|}{%
\begin{tabular}
[c]{|l|}\hline
$8\,$\\\hline
$3$\\\hline
\end{tabular}
} & \multicolumn{1}{|r|}{%
\begin{tabular}
[c]{|l|}\hline
$39\,$\\\hline
\multicolumn{1}{|r|}{$9$}\\\hline
\end{tabular}
} & \multicolumn{1}{|r|}{%
\begin{tabular}
[c]{|l|}\hline
$226\,$\\\hline
\multicolumn{1}{|r|}{$34$}\\\hline
\end{tabular}
} & \multicolumn{1}{|r|}{%
\begin{tabular}
[c]{|l|}\hline
$1,538\,$\\\hline
\multicolumn{1}{|r|}{$135$}\\\hline
\end{tabular}
}\\\hline%
\begin{tabular}
[c]{|l|}\hline
\#$\mathcal{G}_{S,R}$\\\hline
\#r\\\hline
\#{\small pss}\\\hline
\end{tabular}
& \multicolumn{1}{|r|}{%
\begin{tabular}
[c]{|l|}\hline
$1$\\\hline
$1$\\\hline
$1$\\\hline
\end{tabular}
} & \multicolumn{1}{|r|}{%
\begin{tabular}
[c]{|l|}\hline
$8$\\\hline
$9$\\\hline
$1$\\\hline
\end{tabular}
} & \multicolumn{1}{|r|}{%
\begin{tabular}
[c]{|r|}\hline
$48$\\\hline
\multicolumn{1}{|l|}{$124$}\\\hline
$4$\\\hline
\end{tabular}
} & \multicolumn{1}{|r|}{%
\begin{tabular}
[c]{|r|}\hline
$299$\\\hline
\multicolumn{1}{|l|}{$1,653$}\\\hline
$7$\\\hline
\end{tabular}
} & \multicolumn{1}{|r|}{%
\begin{tabular}
[c]{|r|}\hline
$2,059\,$\\\hline
\multicolumn{1}{|l|}{$25,512$}\\\hline
$23$\\\hline
\end{tabular}
}\\\hline%
\begin{tabular}
[c]{|l|}\hline
\#$\mathcal{G}_{S,R,\text{red}}$\\\hline
\#r\\\hline
\#{\small pss}\\\hline
\end{tabular}
& \multicolumn{1}{|r|}{%
\begin{tabular}
[c]{|l|}\hline
$0$\\\hline
$0$\\\hline
$0$\\\hline
\end{tabular}
} & \multicolumn{1}{|r|}{%
\begin{tabular}
[c]{|l|}\hline
$5$\\\hline
$6$\\\hline
$1$\\\hline
\end{tabular}
} & \multicolumn{1}{|r|}{%
\begin{tabular}
[c]{|l|}\hline
$32$\\\hline
$92$\\\hline
\multicolumn{1}{|r|}{$1$}\\\hline
\end{tabular}
} & \multicolumn{1}{|r|}{%
\begin{tabular}
[c]{|r|}\hline
$204$\\\hline
\multicolumn{1}{|l|}{$1,295$}\\\hline
$6$\\\hline
\end{tabular}
} & \multicolumn{1}{|r|}{%
\begin{tabular}
[c]{|r|}\hline
$1,465$\\\hline
\multicolumn{1}{|l|}{$20,680$}\\\hline
$12$\\\hline
\end{tabular}
}\\\hline
\end{tabular}
\ \ \ \ \ \ \label{cp_table1}%
\end{equation}
This table gives for each order the number of semigroups which allows to generate: S-expanded algebras $\mathcal{G}_{S}$, reduced algebras $\mathcal{G}_{S,\text{red}}$, resonant subalgebras $\mathcal{G}_{S,R}$ and reductions of the resonant subalgebras $\mathcal{G}_{S,R,\text{red}}$.
The number of expansions preserving semisimplicity, denoted by \#pss, is given for each case. Besides, as shown in Sections \ref{gen_checkings} and \ref{gen_checkings_all}, a given semigroup might have more than one resonance and thus, for the cases with resonances, we also give the total number of resonances \#r.

The fraction of semigroups preserving semisimplicity is very small and remarkably, as shown explicitly in Section 4.2 of Ref. \cite{Andrianopoli:2013ooa}, expansions performed with those semisgroups always contains the original algebra as a subalgebra. 
For example, the semigroups $S_{\left(  n\right)}^{a}$ of order $n=3,4,5$ with a zero element and a resonant decomposition that preserve semisimplicity are: 
\[
S_{\left(  3\right)  }^{12},\ S_{\left(  4\right)  }^{88},\ S_{\left(
5\right)  }^{770},\ S_{\left(  5\right)  }^{779},\ S_{\left(  5\right)
}^{922},\ S_{\left(  5\right)  }^{968},\ S_{\left(  5\right)  }^{990}%
,\ S_{\left(  5\right)  }^{991}\,.
\]
For some of them it is possible to perform more than one reduction consecutively and thus, it can be shown that the semigroups $S_{\left(  3\right)  }^{12}$, $S_{\left(  4\right) }^{88}$, $S_{\left(  5\right)  }^{779}$ and $S_{\left(  5\right)  }^{922}$ leads after the whole process to the trivial result $\mathfrak{sl}(2,\mathbb{R})$.
Expansions with $S_{\left(  5\right)  }^{770}$, $S_{\left(  5\right)  }^{968}$, $S_{\left(5\right)  }^{990}$ leads essentially to $\mathfrak{sl}(2,\mathbb{R})\oplus\mathfrak{sl}(2,\mathbb{R})$, as it can be shown by performing a suitable change of basis. Finally, the expansion the semigroup $S_{\left(  5\right)  }^{991}$ leads to $\mathfrak{su}\left(  2,\mathbb{R}\right)
\oplus\mathfrak{sl}\left(  2,\mathbb{R}\right)  $.

One of the motivations used in \cite{Andrianopoli:2013ooa} to study S-expansions preserving semisimplicity was the possibility of establishing some relations between the classical simple Lie algebras $A_{n}$, $B_{n}$, $C_{n}$, $D_{n}$ and the special ones. The detailed analysis revealed that the expansions preserving semisimplicity generate only direct sums of simple algebras, always containing the original one as a subalgebra\footnote{This is consistent with the result obtained recently in Ref. \cite{Artebani:2016gwh}, were it was stated that the expansion of simple algebras leads in general to non-simple algebras.}. 
This lead to the conjecture that there is no S-expansion procedure relating them (at least with the resonant conditions considered here). However, a general proof for this statement still remains as an open problem. 

Finally, the programs 36-41 are examples to study the compactness property of S-expanded alegbras (with semigroups of order $n=2,3,4$), in terms of the eigenvalues of the semigroup metric and the expanded Killing-Cartan metric given respectively in Eqs. (\ref{gs}) and (\ref{KCexp}). Following the procedure described in the previous section, these programs can be extended in order to study the compactness of resonant subalgebras and reduced algebras.

\section{Comments}
\label{comments}

Motivated by the growing number of applications that the S-expansion method have had recently, mainly in the construction of higher dimensional gravity theories and the understanding of their interrelations, we have developed a computational tool to automatize this procedure. It is given as a Java library which is composed by 11 classes (listed in Section \ref{descriptionLib} and described in Sections \ref{generating} and \ref{Library_Sexpansion}) whose methods allows:
\begin{itemize}
	\item To perform basic operations with semigroups (check associativity and commutativity, find the zero element, resonances and isomorphisms),
	\item To represent arbitrary Lie algebras and semigroups in order to perform S-expansions with arbitrary semigroups.
\end{itemize}
Many examples has been provided in Section \ref{other_app} to solve a considerable number of problems. They are presented such that any user, not necessarilly an expert in Java, can easily modify them to perform his own calculations.

An important input we have used in many of our algorithms is the lists of all the non isomorphic semigroups 
generated by the Fortran program \textit{gen.f} given in Ref. \cite{Hildebrant}. Consequently, this is an input based in the results of many works related to the classification of semigroups (some of them, the most importants for our purposes, were cited in Section \ref{hist_semig}). 
As mentioned in Ref. \cite{Hildebrant}, the program \textit{gen.f} should be able to generate in principle the full lists up to order 8. However, in our case we got them only up to order 6, because for some reason the program stops when reaches the 835,927th of the 836,021 semigroups of order 7. 
That is why, when we need to have the list of all non-isomorphic semigroups of a give order, we resctrict our calculations up to order 6. 
However, with our library we can still perform calculations with semigroups of order higher than 6 (see e.g., the examples in Sections \ref{asso_com} and \ref{gen_checkings2}).
The only issue is that we do not have the full list of non isomorphic tables for those higher orders.

Besides, as our library has an open licence GNU\footnote{As mentioned before anyone can download, use and modify the library. The corresponding citation to this original version is appreciated.}, we expect not only that these lists can be updated soon for semigroups up to order 8 but also that some of its methods can be improved and extended. For example, a superalgebra has usually a subspace structure $\mathcal{G}$ $=V_{0}\oplus V_{1}\oplus V_{2}$ given by
\begin{align*}
\left[  V_{0},V_{0}\right]    & \subset V_{0}\ ,\ \ \left[  V_{0}%
,V_{1}\right]  \subset V_{1}\ ,\ \ \left[  V_{0},V_{2}\right]  \subset V_{2}\\
\left[  V_{1},V_{1}\right]    & \subset V_{0}\oplus V_{2}\ ,\ \ \left[
V_{1},V_{2}\right]  \subset V_{1}\ ,\ \ \left[  V_{2},V_{2}\right]  \subset
V_{0}\oplus V_{2}%
\end{align*}
so, according with Ref. \cite{irs}, in order to perform S-expansions of superalgebras one needs to consider resonant decompositions $S=S_{0}\cup S_{1}\cup S_{2}$ satisfying
\begin{align*}
S_{0}\times S_{0}  & \subset S_{0}\ ,\ \ S_{0}\times S_{1}\subset
S_{1}\ ,\ \ S_{0}\times S_{2}\subset S_{2}\\
S_{1}\times S_{1}  & \subset S_{0}\cap S_{2}\ ,\ \ S_{1}\times S_{2}\subset
S_{1}\ ,\ \ S_{2}\times S_{2}\subset S_{0}\cap S_{2}%
\end{align*}
Therefore, an extension of the method described in Section \ref{Resonant} for this kind of resonances would allow to perform S-expansions of superalgebras with all the machinery developed in this work. It would also be useful to extend the library to perform \textit{resonant reductions} which, as briefly mentioned at the end of Section \ref{sexpp}, are more general than the $0_{S}$-reduction process.

Interestingly, in Ref. \cite{Ipinza:2016bfc} it has been recently proposed an analytic method to answer if two given Lie algebras can be S-related. They consider resonances in which an element, different than zero, is not allowed to be repeated in the subsets of the considered decomposition of the semigroup. 
As our methods are able to find all possible resonant decompositions (allowing non-zero elements to be repeated), a method selecting the decompositions with non-repeated elements may be implemented in such a way that our library can also be useful for this kind of problem, at least for the case $S=S_{0}\cup S_{1}$. 
We will provide those methods in Ref. \cite{Inostroza} as a first extension of this library, together with an independent and complementary procedure that we developed in parallel, in order to answer whether two given Lie algebras can be S-related.

On the other hand, as mentioned in the introduction, in the early 60's it was conjectured that all solvable Lie algebras could be obtained as a contraction of a semisimple Lie algebra of the same dimension (see, e.g., \cite{Zaitsev,Celeghini}). Although in 2006 there were found some contra-examples that disproved this supposition \cite{Nesterenko} that problem has been revisited very recently in the context of the S-expansion \cite{Nesterenko2012}. Whether this method fixes the classification of solvable Lie algebras still remains an open problem, therefore we think that the computational tools developed in this work might be useful to work this out.

It is also worth to mention that the fact that the S-expansion reproduces all the WW contractions does not imply that the S-expansion covers all possible contractions.
As noticed in \cite{Nesterenko2012} there exist contractions that cannnot be realized by a WW contraction (see e.g., Refs.\cite{Burde,Popovych}) and that is why it is still unclear whether all contractions can be obtained from S-expansions. 
On the other hand, we do know that there exist S-expansions that are not equivalent to any contraction (an explicit example is given in \cite{Nesterenko2012}) and, in general, this is the case of S-expansions with semigroups that do not belong to the family $S_{E}^{\left(  N\right)}$ and/or the case of expansions that lead to algebras of dimension bigger than the original one. Thus, the computational tool provided on this article might also be useful to answer if S-expansions exhaust or not all possible contractions.

From the theoretical-physics point of view, the S-expansion method has been used to show that Chern-Simons (CS) or Born Infeld (BI) type gravities constructed with the $S_{E}^{\left(N\right)}$ family lead, on a certain limit, to higher dimensional General Relativity. Thus, between the possible new applications, the techniques developed in this paper might help to analyze if it is possible to uncover and classify all the relations of this type (some results on this particular problem will be submitted soon \cite{CS_GR}).
Besides, an extension of the method to calculate resonant decompositions able to expand superalgebras might also be very useful in the context of supergravity (see e.g., \cite{Arnowitt:1975xg,Akulov:1975ax}). In particular, it might be analyzed whether is possible to obtain standard five-dimensional supergravity (see \cite{Dauria}) from a Chern-Simons supergravity based on a suitable expanded superalgebra.

Another application is related with the so called Pure Lovelock (PL) theory \cite{Cai:2006pq,Dadhich:2012ma,Dadhich:2015ivt}, which is a higher dimensional theory that has recently called the attention by the fact that their black hole solutions are asymptotically indistinguishable from the ones in general relativity. In Refs. \cite{Concha:2016kdz,Concha:2016tms} an atempt to obtain PL gravity from CS and BI type gravities has been carried out using the expanded algebras. The relation is obtained at the level of the action, but present some problems at the level of the field equations, which can be fixed only by performing suitable identifications of the fields. Thus, the methods of this library could be used to look for a suitable S-expanded algebra giving the right dynamical limit and without performing identification of the fields (see \cite{CS_PL_Gravity}).

Apart from the mentioned applications in gravity theories and Lie group theory, it would be interesting analize if this tool could have also applications in other physical theories related with symmetries as well as in other branches, like semigroup theory and constraint logic programming. Besides, Lie algebras and semigroups has been recently applied in robotics and artificial intelligence so it would also be interesting to explore possible applications of this methods in this area too.

\section*{Acknowledgements}

C.I. was supported by a Mecesup PhD grant and the Término de tesis grant from CONICYT (Chile).
I.K. was supported by was supported by Fondecyt (Chile) grant 1050512 and by DIUBB (Chile) Grant Nos.  102609 and  GI 153209/C. 
N.M. was supported by the FONDECYT (Chile) grant 3130445 during the first stage of this work, and then by a Becas-Chile postdoctoral grant. 
F. N. wants to thank CSIC for a JAE-Predoc grant cofunded by the European Social Fund. 

\appendix

\section{List of program examples}
\label{list_ex}

In this Appedix we give the list of 45 example programs included in the file \textit{examples.zip} available in \cite{webJava}. Most of them has been described in Section \ref{other_app} and, according to the classification explained in section \ref{descriptionLib}, they are called generically as ``prefix\_name''.

\begin{small}
\begin{enumerate}
	\item I\_associative\_checking.java
	\item I\_commutative\_checking.java
	\item I\_commutative\_ord2\_to\_6.java
	\item I\_isomorphisms\_ex1.java
	\item I\_isomorphisms\_ex2.java
	\item I\_isomorphisms\_ex3.java	
	\item I\_isomorphisms\_SE\_N.java
	\item I\_Permutations\_and\_inverses\_console\_n4.java
	\item I\_Permutations\_and\_inverses\_Order\_2\_to\_7.java
	\item I\_PermuteWith\_ex.java
	\item II\_AdjointRep\_Casimirs\_ex.java
	\item II\_Example\_diagonalization.java
	\item II\_findAllResonances\_console\_ord4.java
	\item II\_findAllResonances\_ord2\_to\_6.java
	\item II\_findresonances\_ex\_console.java
	\item II\_findresonances\_ex.java
	\item II\_findzero\_and\_AllResonances\_ord2\_to\_6.java
	\item II\_findzero\_console\_ord4.java
	\item II\_findzero\_ex.java
	\item II\_findzero\_ord2\_to\_6.java
	\item II\_isresonant\_ex.java
	\item II\_isresonant\_permute\_ex1.java
	\item II\_isresonant\_permute\_ex2.java
	\item II\_isresonant\_permute\_ex3.java
	\item II\_isresonant\_permute\_ex4.java
	\item II\_S\_and\_G\_AdjointRep\_ex.java
	\item II\_SExp\_CheckingByHand.java
	\item II\_SExp\_sl2\_S770.java
	\item II\_SExp\_sl2\_S968.java
	\item II\_SExp\_sl2\_S990.java
	\item II\_SExp\_sl2\_S991.java
	\item II\_SExpStructConst\_sl2\_S770.java
	\item II\_SExpStructConst\_sl2\_S968.java
	\item II\_SExpStructConst\_sl2\_S990.java
	\item II\_SExpStructConst\_sl2\_S991.java
	\item III\_EigenVectors\_SExp\_sl2\_ord2.java
	\item III\_EigenVectors\_SExp\_sl2\_ord3.java
	\item III\_EigenVectors\_SExp\_sl2\_ord4.java
	\item III\_Signature\_Sem\_ord2.java
	\item III\_Signature\_Sem\_ord3.java
	\item III\_Signature\_Sem\_ord4.java
	\item III\_sl2\_SExp\_ord2\_to\_6\_console.java
	\item III\_sl2\_SExp\_Red\_ord2\_to\_6\_console.java
	\item III\_sl2\_SExp\_Res\_ord2\_to\_6\_console.java
	\item III\_sl2\_SExp\_Res\_Red\_ord2\_to\_6\_console.java

\end{enumerate}
\end{small}


\begin{thebibliography}{99}                                                                                               %


\bibitem {Segal} I.E. Segal, ``A class of operator algebras which are determined by groups,'' Duke Math. J
18(1951) 221

\bibitem {IW} E. In\"{o}n\"{u} and E.P. Wigner, ``On the contraction of groups
and their representations,'' Proc. Nat. Acad. Sci. U.S.A. 39, 510-524 (1953);
E. In\"{o}n\"{u}, ``Contractions of Lie groups and their representations,'' in
``Group theoretical concepts in elementary particle physics,'' F.
G\"{u}rseyed., Gordon and Breach, pp. 391-402 (1964)

\bibitem{Saletan} E.J. Saletan, ``Contractions of Lie groups,'' J.Math.
Phys. \textbf{2}, 1-21 (1961).


\bibitem{def1} M. Gerstenhaber, ``On the deformations of rings and
algebras,'' Ann. Math. \textbf{79}, 59-103 (1964)

\bibitem{def2} A. Nijenhuis and R.W. Richardson Jr., ``Cohomology and
deformations in graded Lie algebras,'' Bull. A, Math. Soc. \textbf{72}, 1-29
(1966); A. Nijenhuis and R.W. Richardson Jr., \textit{Deformations of
Lie algebra structures}, J. Math. Mech. \textbf{171}, 89-105 (1967)

\bibitem{def3} R.W. Richardson, ``On the rigidity of semi-direct
products of Lie algebras,'' Pac. J. Math. \textbf{22}, 339-344 (1967)


\bibitem {Barut} Barut A. O. and Ratzka R. 1986 \textit{Theory of Group Representations and Applications} (Singapore:World Scientific)


\bibitem {Gilmore} Gilmore R. 1974 \textit{Lie Groups, Lie Algebras, and Some of their Applications} (New York: Wiley-Interscience)


\bibitem {WW}E. Weimar-Woods, ``Contractions of Lie algebras: generalized
Inonu-Wigner contractions versus graded contractions,'' J. Math. Phys. 36,
4519-4548 (1995); Jour. Math. Phys. 32 (1991) 2028; E. Weimar-Woods,
``Contractions, generalized In\"{o}n\"{u} and Wigner contractions and
deformations of finite-dimensional Lie algebras,'' Rev. Math. Phys. 12,
1505-1529 (2000)


\bibitem {hs}M.~Hatsuda and M.~Sakaguchi, ``Wess-Zumino term for the AdS
superstring and generalized Inonu-Wigner  contraction,''
Prog.\ Theor.\ Phys.\ \textbf{109} (2003) 853 [arXiv:hep-th/0106114].



\bibitem {aipv1}J.~A.~de Azcarraga, J.~M.~Izquierdo, M.~Picon and O.~Varela,
``Generating Lie and gauge free differential (super)algebras by expanding
Maurer-Cartan forms and Chern-Simons supergravity,'' Nucl.\ Phys.\ B
\textbf{662} (2003) 185 [arXiv:hep-th/0212347].


\bibitem {aipv2}J.~A.~de Azcarraga, J.~M.~Izquierdo, M.~Picon and O.~Varela,
``Extensions, expansions, Lie algebra cohomology and enlarged superspaces,''
Class.\ Quant.\ Grav.\ \textbf{21} (2004) S1375  [arXiv:hep-th/0401033].


\bibitem {aipv3}J.~A.~de Azcarraga, J.~M.~Izquierdo, M.~Picon and O.~Varela,
``Expansions of algebras and superalgebras and some applications,''
Int.\ J.\ Theor.\ Phys.\ \textbf{46} (2007) 2738  [arXiv:hep-th/0703017]. 

\bibitem {Nakahara} M. Nakahara, \textit{Geometry, topology and physics}, Institute of Physics
publishing (1990).


\bibitem {irs}F.~Izaurieta, E.~Rodriguez and P.~Salgado,  ``Expanding Lie
(super)algebras through Abelian semigroups,''  J.\ Math.\ Phys.\ \textbf{47}
(2006) 123512  [arXiv:hep-th/0606215].


\bibitem {Diaz:2012zza}J.~D\'{\i}az, O.~Fierro, F.~Izaurieta, N.~Merino,
E.~Rodr\'{\i}guez, P.~Salgado and O.~Valdivia, ``A generalized action for (2 +
1)-dimensional Chern-Simons gravity,'' J.\ Phys.\ A \textbf{45}, 255207
(2012), [arXiv:1311.2215 [gr-qc]].

\bibitem{Zaitsev} Zaitsev G.A., ``Group-invariant study of the sets of limiting geometrie and special Lie subalgebras,'' 1961, Talk thesises of All-USSR Geometric Conference (Kiev, USSR, 1961), 1-48 (Russian).

\bibitem{Celeghini} Celeghini E. and Tarlini M., ``Contraction of group representations II,'' Nuovo Cimento B 65 (1981), 172-180.

\bibitem{Nesterenko} M. Nesterenko, R. Popovych, ``Contractions of low-dimensional Lie algebras,''  
J.\ Math.\ Phys.\  {\bf 47} (2006) 123515.


\bibitem{Nesterenko2012} Maryna Nesterenko, ``S-expansions of three-dimensional Lie algebras,'' Institute of Mathematics of NAS of Ukraine, 3 Tereshchenkivs'ka Str., Kyiv-4, 01601 Ukraine, arXiv:1212.1820; Maryna Nesterenko, ``S-expansions of three-dimensional Lie algebras,'' Group analysis of differential equations and integrable systems, 147-154, Department of
Mathematics and Statistics, University of Cyprus, Nicosia, 2013.

\bibitem {Zanelli:2005sa} J.~Zanelli,  ``Lecture notes on Chern-Simons
(super-)gravities. Second edition (February 2008),''  hep-th/0502193.

\bibitem{Izaurieta:2006aj} 
  F.~Izaurieta, E.~Rodriguez and P.~Salgado, ``Eleven-dimensional gauge theory for the M algebra as an Abelian semigroup expansion of osp(32|1),''
  Eur.\ Phys.\ J.\ C {\bf 54}, 675 (2008), [hep-th/0606225].


\bibitem {irs2}F.~Izaurieta, E.~Rodriguez, A.~Perez and P.~Salgado,  ``Dual
Formulation of the Lie Algebra S-expansion Procedure,''
J.\ Math.\ Phys.\ \textbf{50} (2009) 073511  [arXiv:0903.4712 [hep-th]].


\bibitem {Edelstein:2006se} J.~D.~Edelstein, M.~Hassaine, R.~Troncoso and
J.~Zanelli,  ``Lie-algebra expansions, Chern-Simons theories and the
Einstein-Hilbert Lagrangian,''  Phys.\ Lett.\ B \textbf{640}, 278 (2006)
[hep-th/0605174].


\bibitem {Izaurieta:2009hz} F.~Izaurieta, E.~Rodriguez, P.~Minning, P.~Salgado
and A.~Perez,  ``Standard General Relativity from Chern-Simons Gravity,''
Phys.\ Lett.\ B \textbf{678}, 213 (2009)  [arXiv:0905.2187 [hep-th]].


\bibitem {Concha:2013uhq} P.~K.~Concha, D.~M.~Peñafiel, E.~K.~Rodr\'iguez and
P.~Salgado,  ``Even-dimensional General Relativity from Born-Infeld
gravity,''  Phys.\ Lett.\ B \textbf{725}, 419 (2013)  [arXiv:1309.0062
[hep-th]].

\bibitem {Concha:2014vka} P.~K.~Concha, D.~M.~Peñafiel, E.~K.~Rodr\'iguez and
P.~Salgado,  ``Chern-Simons and Born-Infeld gravity theories and Maxwell
algebras type,''  Eur.\ Phys.\ J.\ C \textbf{74}, 2741 (2014)
[arXiv:1402.0023 [hep-th]].

\bibitem{Concha:2014zsa} P.~K.~Concha, D.~M.~Peñafiel, E.~K.~Rodr\'iguez and P.~Salgado, ``Generalized Poincar\'e algebras and Lovelock-Cartan gravity theory,'' 
Phys.\ Lett.\ B {\bf 742}, 310 (2015) [arXiv:1405.7078 [hep-th]].


\bibitem {DG}S. Deser, G.W. Gibbons, ``Born-Infeld-Einstein Actions?,''
Class. Quant. Grav. \textbf{15} (1998) L35, [hep-th/9803049].

\bibitem{Quinzacara:2012zz} C.~A.~C.~Quinzacara and P.~Salgado, ``Black hole for the Einstein-Chern-Simons gravity,'' Phys.\ Rev.\ D {\bf 85}, 124026 (2012) [arXiv:1401.1797 [gr-qc]].
	
\bibitem{Quinzacara:2013uua} C.~A.~C.~Quinzacara and P.~Salgado, ``Stellar equilibrium in Einstein-Chern-Simons gravity,'' Eur.\ Phys.\ J.\ C {\bf 73}, no. 6, 2479 (2013).

\bibitem{Crisostomo:2014hia} J.~Crisóstomo, F.~G\'omez, P.~Salgado, C.~Quinzacara, M.~Cataldo and S.~del Campo, ``Accelerated FRW Solutions in Chern-Simons Gravity,''  Eur.\ Phys.\ J.\ C {\bf 74}, no. 10, 3087 (2014) [arXiv:1401.2128 [gr-qc]].

\bibitem{Crisostomo:2016how} J.~Crisóstomo, F.~G\'omez, C.~Quinzacara and P.~Salgado, ``Static solutions in Einstein-Chern-Simons gravity,'' arXiv:1601.06592 [gr-qc].

\bibitem{Gonzalez:2016xwo} N.~González, G.~Rubio, P.~Salgado and S.~Salgado, ``Generalized Galilean algebras and Newtonian gravity,'' Phys.\ Lett.\ B {\bf 755}, 433 (2016) [arXiv:1604.06313 [hep-th]].


\bibitem{Caroca:2010ax} R.~Caroca, N.~Merino and P.~Salgado, ``S-Expansion of Higher-Order Lie Algebras,'' J.\ Math.\ Phys.\  {\bf 50}, 013503 (2009) [arXiv:1004.5213 [math-ph]].

\bibitem{Caroca:2010kr} R.~Caroca, N.~Merino, A.~Perez and P.~Salgado, ``Generating Higher-Order Lie Algebras by Expanding Maurer Cartan Forms,'' J.\ Math.\ Phys.\  {\bf 50}, 123527 (2009) [arXiv:1004.5503 [hep-th]].

\bibitem{Caroca:2011zz} R.~Caroca, N.~Merino, P.~Salgado and O.~Valdivia, ``Generating infinite-dimensional algebras from loop algebras by expanding Maurer-Cartan forms,'' J.\ Math.\ Phys.\  {\bf 52}, 043519 (2011).


\bibitem {Caroca:2011qs} R.~Caroca, I.~Kondrashuk, N.~Merino and F.~Nadal,
``Bianchi spaces and their three-dimensional isometries as S-expansions of
two-dimensional isometries,''  J.\ Phys.\ A \textbf{46}, 225201 (2013)
[arXiv:1104.3541 [math-ph]].


\bibitem {bian} L. Bianchi, ``Sugli spazi a tre dimensioni che ammettono un
gruppo continuo di movimenti,'' Memorie \ di Matematica e di Fisica della
Societa Italiana delle Scienze, Serie Terza, Tomo XI, pp. 267--352 (1898).


\bibitem {Soroka:2006aj}D.~V.~Soroka and V.~A.~Soroka, \textquotedblleft
Semi-simple extension of the (super)Poincare algebra\textquotedblright%
,\ Adv.\ High Energy Phys.\ \textbf{2009}, 234147 (2009), [hep-th/0605251].


\bibitem{Durka:2011nf} R.~Durka, J.~Kowalski-Glikman and M.~Szczachor, ``Gauged AdS-Maxwell algebra and gravity,'' Mod.\ Phys.\ Lett.\ A {\bf 26}, 2689 (2011) [arXiv:1107.4728 [hep-th]].

\bibitem{Durka:2011gm} R.~Durka, J.~Kowalski-Glikman and M.~Szczachor, ``AdS-Maxwell superalgebra and supergravity,'' Mod.\ Phys.\ Lett.\ A {\bf 27}, 1250023 (2012) [arXiv:1107.5731 [hep-th]].


\bibitem {Salgado:2014qqa}P.~Salgado and S.~Salgado, ``$\mathfrak{so}%
(D-1,1)\otimes\mathfrak{so}(D-1,2)$ algebras and gravity,'' Phys.\ Lett.\ B
\textbf{728}, 5 (2014).

\bibitem{Concha:2016hbt} P.~K.~Concha, R.~Durka, N.~Merino and E.~K.~Rodr\'iguez, ``New family of Maxwell like algebras,'' Phys.\ Lett.\ B {\bf 759}, 507 (2016), [arXiv:1601.06443 [hep-th]].

\bibitem{Salgado:2014jka} P.~Salgado, R.~J.~Szabo and O.~Valdivia, ``Topological gravity and transgression holography,'' Phys.\ Rev.\ D {\bf 89}, no. 8, 084077 (2014) [arXiv:1401.3653 [hep-th]].

\bibitem{Fierro:2014lka} O.~Fierro, F.~Izaurieta, P.~Salgado and O.~Valdivia, ``(2+1)-dimensional supergravity invariant under the AdS-Lorentz superalgebra,'' arXiv:1401.3697 [hep-th].

\bibitem{Concha:2014xfa} 
  P.~K.~Concha and E.~K.~Rodr\'iguez, ``Maxwell Superalgebras and Abelian Semigroup Expansion,''
  Nucl.\ Phys.\ B {\bf 886}, 1128 (2014), [arXiv:1405.1334 [hep-th]].

\bibitem{Concha:2014tca} 
  P.~K.~Concha and E.~K.~Rodr\'iguez, ``N = 1 Supergravity and Maxwell superalgebras,''
  JHEP {\bf 1409}, 090 (2014), [arXiv:1407.4635 [hep-th]].   

\bibitem{Concha:2015tla} 
  P.~K.~Concha, E.~K.~Rodr\'iguez and P.~Salgado, ``Generalized supersymmetric cosmological term in $N=1$ Supergravity,''
  JHEP {\bf 1508}, 009 (2015), [arXiv:1504.01898 [hep-th]].

\bibitem{Concha:2015woa} P.~K.~Concha, O.~Fierro, E.~K.~Rodr\'iguez and P.~Salgado, ``Chern-Simons supergravity in $D=3$ and Maxwell superalgebra,''
  Phys.\ Lett.\ B {\bf 750}, 117 (2015), [arXiv:1507.02335 [hep-th]].


\bibitem{Ipinza:2016con} M.~C.~Ipinza, P.~K.~Concha, L.~Ravera and E.~K.~Rodr\'iguez, ``On the Supersymmetric Extension of Gauss-Bonnet like Gravity,'' JHEP {\bf 1609}, 007 (2016) [arXiv:1607.00373 [hep-th]].

\bibitem{Concha:2016zdb} P.~K.~Concha, O.~Fierro and E.~K.~Rodr\'iguez, ``In\"on\"u-Wigner Contraction and $D=2+1$ Supergravity,'' arXiv:1611.05018 [hep-th].


\bibitem{Durka:2016eun} R.~Durka, ``Resonant algebras and gravity,'' arXiv:1605.00059 [hep-th].

\bibitem{Penafiel:2016ufo} D.~M.~Peñafiel and L.~Ravera, ``Generalized In\"on\"u-Wigner Contraction as $S$-Expansion with Infinite Semigroup and Ideal Subtraction,'' arXiv:1611.05812 [hep-th].


\bibitem {Andrianopoli:2013ooa} L.~Andrianopoli, N.~Merino, F.~Nadal and
M.~Trigiante,  ``General properties of the expansion methods of Lie
algebras,''  J.\ Phys.\ A \textbf{46}, 365204 (2013)  [arXiv:1308.4832
[gr-qc]].


\bibitem {webJava} https://github.com/SemigroupExp/Sexpansion/releases/tag/v1.0.0


\bibitem {n6-2}R. Plemmons, ``A survey of computer applications to semigroups
and related structures,'' ACM SIGSAM 12, (1969) 28-39.

\bibitem {irs_IZ} F.~Izaurieta, ``Semigroup Expansion and M-Supergravity in Eleven Dimensions,'' hep-th/0611238.

\bibitem{Gonzalez:2014tta} N.~Gonz\'alez, P.~Salgado, G.~Rubio and S.~Salgado, ``Einstein-Hilbert action with cosmological term from Chern-Simons gravity,'' J.\ Geom.\ Phys.\  {\bf 86}, 339 (2014).

\bibitem {n4}G.E. Forsythe, ``SWAC computes 126 distinct semigroups of order
4,'' Proc. Amer. Math. Soc. 6, (1955) 443-447

\bibitem {n5}T.S. Motzkin and J.L. Selfridge, ``Semigroups of order five,''
The November meeting in Los Angeles, Bull. Amer. Math. Soc. 62 (1) (1956) 13-23

\bibitem {n6-1}R. Plemmons, ``Construction and analysis of non-equivalent
finite semigroups'' (1970) Computational Problems in Abstract Algebra (Proc.
Conf., Oxford, 1967) pp. 223-228 Pergamon, Oxford

\bibitem {n6-3}R. Plemmons, ``There are 15973 semigroups of order 6,'' Math.
Algorithms 2, (1967) 2-17

\bibitem {n7}H. J\"urgensen, P. Wick, ``Die Halbgruppen der Ordnungen $\leq$
7,'' Semigroup Forum 14 (1977), 1, 69-79.

\bibitem {n8}S. Satoh, K. Yama, M. Tokizawa, ``Semigroups of order 8,''
Semigroup Forum 49 (1994), 1, 7-29.

\bibitem {n9-1}A. Distler, T. Kelsey, ``The monoids of orders eight, nine \&
ten,'' Ann. Math. Artif. Intell. 56 (2009), 3-21.

\bibitem {n9-2}A. Distler and T. Kelsey, ``The monoids of order eight and
nine,'' In S. Autexier, J. Campbell, J. Rubio, V. Sorge, M. Suzuki, and F.
Wiedijk (Eds), Artificial Intelligence and Symbolic Computation, 8th
International Conference, AISC 2008, Birmingham, July, 2008, Proceedings,
volume 5144 of ``Lecture Notes in Computer Science,'' 61-76, Springer, 2008.

\bibitem {n9-3}A. Distler and J.D. Mitchell, ``Smallsemi - A library of small
semigroups,'' http://www-history.mcs.st-and.ac.uk/jamesm/smallsemi/, Feb 2010.

\bibitem {n9-4}A. Distler, T. Kelsey and J.D. Mitchell, http://www-circa.mcs.st-and.ac.uk/

\bibitem {S10} Andreas Distler, Chris Jefferson, Tom Kelsey and Lars Kotthof, ``The Semigroups of Order 10,'' Principles and Practice of Constraint Programming Volume 7514 of the series Lecture Notes in Computer Science pp 883-899, Springer Berlin Heidelberg. 18th International Conference, CP 2012, Québec City, QC, Canada, October 8-12, 2012. Proceedings. DOI 10.1007/978-3-642-33558-7\_63

\bibitem {Hildebrant}Hildebrant J 2001 \textit{Handbook of Finite Semigroup
Programs} (LSU Mathematics Electronic Preprint Series) preprint 2001-24


\bibitem {jama} http://math.nist.gov/javanumerics/jama/


\bibitem {wiki} https://github.com/SemigroupExp/Sexpansion/wiki


\bibitem{Artebani:2016gwh} M.~Artebani, R.~Caroca, M.~C.~Ipinza, D.~M.~Peñafiel and P.~Salgado, ``Geometrical aspects of the Lie algebra S-expansion procedure,'' J.\ Math.\ Phys.\  {\bf 57}, no. 2, 023516 (2016).


\bibitem{Ipinza:2016bfc} M.~C.~Ipinza, F.~Lingua, D.~M.~Peñafiel and L.~Ravera, ``An Analytic Method for $S$-Expansion involving Resonance and Reduction,'' Fortsch.\ Phys.\  {\bf 64}, no. 11-12, 854 (2016) [arXiv:1609.05042 [hep-th]].

\bibitem{Inostroza} C.~Inostroza, I.~Kondrashuk, N.~Merino, F.~Nadal, ``Algorithm to find S-related Lie algebras,'' Will be submitted soon.


\bibitem{Burde} D. Burde, ``Degenerations of 7-dimensional nilpotent Lie algebras,'' Comm. Algebra {\bf 33} (2005), 1259.

\bibitem{Popovych} Popovych D.R. and Popovych R.O., ``Lowest dimensional example on non-universality of generalized In\"{o}n\"{u}-Wigner contractions,'' J. Algebra 324 (2010), 2742-2756.


\bibitem{CS_GR} P.~K.~Concha, C.~Inostroza, N.~Merino and E.~K.~Rodr\'iguez, ``Chern-Simons gravities related with General Relativity,'' Will be submitted soon.


\bibitem{Arnowitt:1975xg} R.~L.~Arnowitt, P.~Nath and B.~Zumino, ``Superfield Densities and Action Principle in Curved Superspace,'' Phys.\ Lett.\  {\bf 56B}, 81 (1975).
	
\bibitem{Akulov:1975ax} V.~P.~Akulov, D.~V.~Volkov and V.~A.~Soroka, ``Gauge Fields on Superspaces with Different Holonomy Groups,'' JETP Lett.\  {\bf 22}, 187 (1975) [Pisma Zh.\ Eksp.\ Teor.\ Fiz.\  {\bf 22}, 396 (1975)].


\bibitem{Dauria} L. Castellani, R. D'Auria and P. Fre, Supergravity and Superstrings: A Geometric Perspective, World Scientific, Singapore, 1991.

\bibitem{Cai:2006pq} R.~G.~Cai and N.~Ohta, ``Black Holes in Pure Lovelock Gravities,'' Phys.\ Rev.\ D {\bf 74}, 064001 (2006)

\bibitem{Dadhich:2012ma} N.~Dadhich, J.~M.~Pons and K.~Prabhu, ``On the static Lovelock black holes,'' Gen.\ Rel.\ Grav.\  {\bf 45}, 1131 (2013)

\bibitem{Dadhich:2015ivt} N.~Dadhich, R.~Durka, N.~Merino and O.~Miskovic, ``Dynamical structure of Pure Lovelock gravity,'' Phys.\ Rev.\ D {\bf 93}, no. 6, 064009 (2016) [arXiv:1511.02541 [hep-th]].


\bibitem{Concha:2016kdz} P.~K.~Concha, R.~Durka, C.~Inostroza, N.~Merino and E.~K.~Rodr\'iguez, ``Pure Lovelock gravity and Chern-Simons theory,'' Phys.\ Rev.\ D {\bf 94}, no. 2, 024055 (2016), [arXiv:1603.09424 [hep-th]].

\bibitem{Concha:2016tms} P.~K.~Concha, N.~Merino and E.~K.~Rodr\'iguez, ``Lovelock gravities from Born-Infeld gravity theory,'' arXiv:1606.07083 [hep-th].

\bibitem{CS_PL_Gravity} R.~Durka, C.~Inostroza and N.~Merino, ``Pure Lovelock gravity from Chern-Simons theory,'' Work in progress.


\end{thebibliography}
\end{document}